\documentclass[%
 reprint,
 amsmath,amssymb,
 aps,
]{revtex4-2}

\usepackage{graphicx}
\usepackage{dcolumn}
\usepackage{bm}
\usepackage{xcolor}

\begin{document}

\preprint{APS/123-QED}
	
	\title{Uncovering the Varieties of Three-dimensional Hall-MHD Turbulence}
	\author{Pratik Patel$^1$} 

       \email[]{d21ph012@phy.svnit.ac.in}
	\author{Sharad K Yadav$^1$}%
	\email{sharadyadav@phy.svnit.ac.in}
	\affiliation{$^1$Department of Physics, Sardar Vallabhbhai National, Institute of Technology (SVNIT), Surat 395007, Gujarat, India}

	\author{Hideaki Miura$^2$}
    \email[]{miura.hideaki@nifs.ac.jp}
	\affiliation{$^2$National Institute for Fusion Science (NIFS), Toki, Gifu 509-5292, Japan}

	\author{Rahul Pandit$^3$}
    \email[]{rahul@iisc.ac.in}
	\affiliation{$^3$Center for Condensed Matter Theory, Department of Physics, Indian Institute of Science (IISc), Banglore 560012, India}
  
	\date{\today}
	
	\begin{abstract}
  We carry out extensive pseudospectral direct numerical simulations (DNSs) of decaying three-dimensional (3D) Hall magnetohydrodynamics (3D HMHD) plasma turbulence at three magnetic Prandtl numbers $Pr_{m}=0.1$, $1.0$ and $10.0$. Our DNSs have been designed to uncover the dependence of the statistical properties of 3D HMHD turbulence on $Pr_m$ and to bring out the subtle interplay between three lengths -- the kinetic and magnetic dissipation length scales $\eta_u$, and $\eta_b$ and the ion-inertial scale $d_i$, below which we see the manifestations of the Hall term. This interplay, qualitatively apparent from isosurface plots of the moduli of the vorticity and the current density, is exposed clearly by the kinetic-energy and magnetic-energy spectra, $E_u(k)$ and $E_b(k)$, respectively. We find two different inertial regions: In the first inertial region $k<k_{i}\sim1/d_i$, both the kinetic-energy and magnetic-energy spectra, $E_u(k)$ and $E_b(k)$, respectively, display power-law regions with an exponent that is consistent with Kolmogorov-type $-5/3$ scaling, at all values of $Pr_m$. In the second inertial region $k > k_{i}$, the scaling of $E_b(k)$  depends upon $Pr_M$: At $Pr_{m}=0.1$, the spectral-scaling exponent is $-17/3$, but for $Pr_{m}=1$  and $10$ this exponent is $-11/3$. We then show theoretically that  
  $E_u(k) \sim k^2 E_b(k)$ for $Pr_m \ll 1$ and $E_b(k) \sim k^2 E_u(k)$ for $Pr_m \gg 1$; our DNS results are consistent with our theoretical predictions. We examine, furthermore, left- and right-polarised fluctuations of the fields that lead, respectively, to the dominance of ion-cyclotron or whistler waves. We also show that 3D HMHD turbulence shows signatures of intermittency that we uncover by analysing the scale dependence of probability distribution functions of velocity- and magnetic-field increments, their structure functions, and their flatnesses as functions of $d_i$ and $Pr_m$.

	\end{abstract}
	
	\maketitle
	
	\section{introduction}\label{sec I}
Eighteen years after the introduction of magnetohydrodynamics (MHD) by Alfv\'en~\cite{alfven1942existence}, Lighthill~\cite{lighthill1960studies} noted that many physical plasmas require the inclusion of the Hall term in the MHD equations. The resulting hydrodynamical partial differential equations (PDEs) are now known as the Hall MHD (HMHD) equations. 
Since their introduction, the HMHD equations have been used in a large variety of physical systems~\cite{Huba2003,shaikh20093d,matthaeus2012review,miura2024formation}, including  magnetic-reconnection processes~\cite{yamada2010magnetic}, pulsars and neutron stars~\cite{shalybkov1997hall,hollerbach2002influence,igoshev2021evolution}, the solar wind~\cite{matthaeus1982measurement,krishan1999astrophysical,matthaeus2012review,kiyani2009global},  the magneto-rotational instability~\cite{o2014multifluid,kopp2022influence},
sub-Alfv\'enic plasma expansion~\cite{ripin1993sub}, and magnetic-field transport in plasma opening switches.

The HMHD equations are used to model plasma processes on length scales shorter than the ion-inertial length $d_i$ and time scales less than the ion-cyclotron period $\sim 1/\omega_{ci}$; here, $d_i\equiv c/\omega_{pi}$, with $\omega_{pi}$ the ion plasma frequency, $\omega_{ci}$ the ion-cyclotron frequency, and $c$ the velocity of light~\cite{krishan2004magnetic,yadav2022statistical,miura2024formation,mahajan2024hall}. Clearly, $d_i$ is an important control parameter here. Indeed, satellite observations of the solar wind~\cite{kiyani2009global} find two inertial regions, with different scaling exponent, in the \textit{magnetic-energy-spectrum} $E_b(k)$. In the first inertial region, the fluid energy spectrum $E_u(k)$ and $E_b(k)$ have a Kolmogorov-type scaling region with a $-5/3$ exponent; the second inertial region in $E_b(k)$ is steeper than the first one and is characterised by a scaling exponent $s \in [-2,-4]$; the crossover from the first to the second region occurs at a wavenumber scale $\sim 1/d_i$. These observations~\cite{kiyani2009global} also uncover intermittency and multiscaling of velocity and magnetic-ﬁeld structure functions  in the first inertial range, but not in the second one.

We examine the effects on HMHD turbulence of another important control parameter, namely, the ratio of the kinematic viscosity $\nu$ to the magnetic diffusivity $\eta$, i.e., the magnetic Prandlt number $Pr_{m}=\frac{\nu}{\eta}=\frac{R_{m}}{R_{e}}$, where $R_{m}$ is the magnetic Reynolds number and $R_{e}$ the kinetic Reynolds number. The value of $Pr_{m}$ ranges over several orders of magnitudes in different plasmas: For the solar corona~\cite{PhysRevLett.92.054502} $Pr_{m} \simeq 10^{-6} - 10^{-4}$; in the interiors of planets and in liquid-metal laboratory dynamos~\cite{RevModPhys.74.973,forest2002spherical,bourgoin2002magnetohydrodynamics} $Pr_{m} \simeq 10^{-5}$;
in the Sun’s convection zone~\cite{RevModPhys.72.1081} $Pr_{m} \simeq 10^{-2}$; for plasma Couette-flow experiments~\cite{ebrahimi2011global} $Pr_{m} \simeq 10^{-3} - 5 \times 10^{2}$; in the interstellar medium (ISM)~\cite{elmegreen2004interstellar,ponty2005numerical,gressel2008dynamo,sahoo2011systematics} the magnetic Prandtl number is very large [$Pr_{m} \simeq 10^{14}$]; it is also very high in the intra-cluster medium (ICM)~\cite{cho2014origin}, in accretion disks, and galactic halos. Different suggestions have been made for the magnetic Prandtl number that should be used for solar-wind turbulence: if the solar wind is considered to be weakly collisional, then both $\nu$ and $\eta$ are small and so is $Pr_{m}$; however, if the solar wind is collisionless, then $\nu$ and $\eta$ must be replaced by effective transport coefficients, which depend on instabilities or turbulence~\cite{howes2011gyrokinetic}, and then $Pr_{m}$ turns out to be large.  

The value of the magnetic Prandtl number has important consequences for dynamo action; in particular, large- and small-scale dynamos are obtained, respectively, at low~\cite{ponty2005numerical} and high~\cite{gressel2008dynamo} $Pr_{m}$. As we will show, $Pr_{m}$ also has significant effects on the statistical properties of HMHD turbulence. Although the $Pr_{m}$-depedence of MHD turbulence has been examined systematically~\cite{sahoo2011systematics}, such a systematic study has not been attempted hitherto. Our work addresses this lacuna.


 In particular, 
we explore spectral scaling regions and intermittency in the three-dimensional (3D) HMHD equations via direct numerical simulations (DNSs), which have been carried out typically at $Pr_{m}=1$; notable exceptions are the DNSs of Refs.~\cite{miura2024formation} for $Pr_{m} > 1$. The DNSs at $Pr_{m}=1$ yield K41-type inertial ranges for both $E_u(k)$ and $E_b(k)$ in the first inertial range; $E_b(k)$ shows a second inertial (or sub-inertial) region, $d_{i} \ll l \ll \eta_{d}^{b}$, where $\eta_{d}^{b}$ is the magnetic-energy dissipation length scale, with either $-7/3$ and $-11/3$ scaling regimes. 
We first show on theoretical grounds that, in the second inertial range, $E_u(k) \sim k^2 E_b(k)$, for $Pr_m \ll 1$ and  $E_b(k) \sim k^2 E_u(k)$, for $Pr_m \gg 1$. We then carry out DNSs that uncover such 
scaling forms. We present a detailed study of the statistical properties of such turbulence at values of $Pr_{m}$ that are different from unity. We compare our results for $Pr_{m}=0.1$, $Pr_{m}=1$, and $Pr_{m}=10$, at  
different values of $d_{i}$, and with $512^3$ and $1024^3$ collocation points in our pseudospectral DNSs. We compute various statistical quantities to characterize 3D HMHD plasma turbulence:
(i) $E_u(k)$ and $E_b(k)$, including spectra conditioned on left- and right-polarised fluctuations of the fields, to uncover ion-cyclotron and whistler waves in the sub-inertial region; (ii) probability distribution functions (PDFs) of longitudinal velocity and magnetic-fields increments, their structure functions, and their flatnesses.  For all values of $Pr_{m}$, there is a K41-type inertial range with  $E_u(k) \sim k^{-5/3}$ and
$E_b(k) \sim k^{-5/3}$. For large $k$, i.e., length scales smaller than $d_i$, we find that $E_b(k) \sim k^{-17/3}$ for $Pr_{m}=0.1$; but $E_b(k) \sim k^{-17/3}$ for  $Pr_{m}=1$ and  $Pr_{m}=10$. We present theoretical arguments for these two different scaling behaviors of $E_b(k)$, in this second-inertial region, and show that they can be understood using left- and right-polarised fluctuations of the fields, which lead to the dominance of either ion-cyclotron or whistler waves. Our analysis of PDFs of field increments, structure functions, and their flatnesses show that the velocity fields are less intermittent than their magnetic-field counterparts for  $Pr_{m}=1$ and  $Pr_{m}=10$.
   The remaining part of this paper is organized as follows. In Section~\ref{sec:Model} we present the incompressible 3D HMHD equations, an outline of our pseudospectral DNS, and the definitions of the statistical quantities we compute to characterize 3D HMHD plasma turbulence. Section~\ref{sec:Results} presents our results and Section~\ref{sec:conc} is devoted to our conclusions and a discussion of our results.
    
    \section{Hall Magnetohydrodynamics and Direct Numerical Simulations}\label{sec:Model}
	Incompressible 3D HMHD is governed by the following set of partial differential equations (PDEs) [see, e.g., Refs.~\cite{lighthill1960studies,krishan1999astrophysical,stawarz2015small,yadav2022statistical,miura2024formation}]:
	\begin{gather}
		\label{eq:vel}
		\frac{\partial\bm{u}}{\partial\textit{t}}+(\bm{u}\cdot\nabla)\bm{u}=-\nabla\bar{\textit{p}}+(\bm{b}\cdot\nabla)\bm{b}+\nu\nabla^2\bm{u}+\bm{f}_u\,;\\
		\label{eq:mag} 
		\frac{\partial\bm{b}}{\partial\textit{t}}=\nabla\times(\bm{u}\times\bm{b}-d_i \bm{j}\times \bm{b})+\eta\nabla^2\bm{b}+\bm{f}_{b}\,;\\ 
		\label{eq:divvel}
		\nabla\cdot\bm{u}=0\,;\\
		\label{eq:divmag}
		\nabla\cdot\bm{b}=0\,; 
	\end{gather}
	here, $\bm{u}$ and $\bm{b}$ are the velocity and magnetic fields, at point $\bm{x}$ and time $t$, the current $\bm{j} = \nabla \times \bm{b}$; $\nu$ and $\eta$ are the  kinematic viscosity and the magnetic diffusivity (or resistivity), respectively, and the magnetic Prandtl number $Pr_{m} \equiv \nu/\eta$; the total pressure $\bar{p}\equiv p+(b^2/8\pi)$, where $p$ is the kinetic pressure and $b^2/8\pi$ the magnetic pressure; ${\bm{f}}_u$ and ${\bm{f}}_b$ are the external forces, which we set to zero because we only consider freely decaying 3D HMHD turbulence. The Hall term in the induction equation~\eqref{eq:mag} has a coefficient $d_i$, which is the ion-inertial length; if $d_i=0$, then HMHD reduces to MHD; Eqs.~\eqref{eq:divvel} and \eqref{eq:divmag} represent the incompressibility condition for the flow and the divergence-free nature of the magnetic field, respectively; the vorticity $\omega=\nabla\times\bm{u}$. 
    
    We solve Eqs.\eqref{eq:vel} - \eqref{eq:divmag} via pseudospectral~\cite{canuto2012spectral,sahoo2011systematics} direct numerical simulations (DNSs) [which yield spectral accuracy] in a 3D, triply periodic, cubical domain with side $L = 2\pi$ and $N^3$ collocation points. We remove aliasing errors using the $2/3$ dealiasing method, so the maximal wave numbers in our DNSs are $k_{max}=171$ and $342$ for $N^{3}=512^{3}$ and $1024^{3}$ collocation points, respectively. In our DNSs, derivatives are evaluated in  Fourier space and products in physical space.
    Such DNSs for 3D HMHD~\cite{yadav2022statistical,miura2014structure,miura2015physics,miura2016hall,miura2019hall} are much more challenging than their 3D MHD counterparts~\cite{sahoo2011systematics} because
    we must have enough spatial resolution to uncover the ion-inertial scale and enough temporal resolution to obtain \textit{whistler waves} and \textit{ion-inertial waves}~\cite{krishan2004magnetic,galtier2006wave,miura2019hall}. For time marching we use the second-order slaved, Adams-Bashforth scheme~\cite{sahoo2011systematics,yadav2022statistical}.
    
\begin{figure}
	\centering
	\includegraphics[width=6.6cm,height=5.3cm]
    {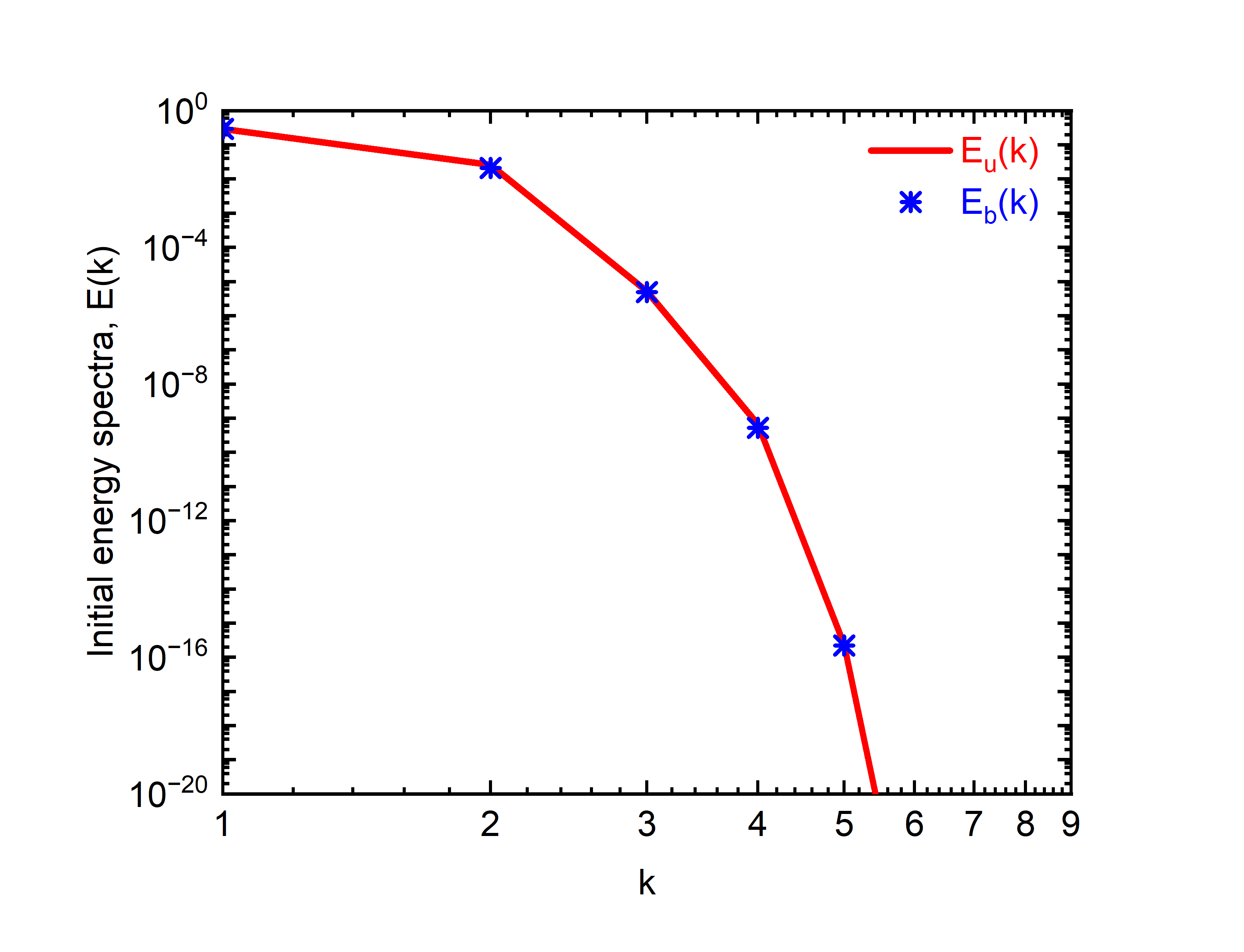}\hfill
		\caption{Log-log plots versus the wave-number $k$ of the initial fluid-energy spectrum~ [red curve] and the magnetic spectrum [shown by stars] given in Eq.~\eqref{eq:initialspectra}.}
		\label{fig:1}
\end{figure}

    We use the following initial $(t=0)$ energy spectra [Fig.~\ref{fig:1}]:
	\begin{equation}
	   E_u^0(k)=E_b^0(k)=E^0k^4\exp(-2k^2);\hspace{0.3cm}E^0=10\,;
	\label{eq:initialspectra}
    \end{equation}
    and we choose the phases of each one of the velocity and magnetic-field Fourier modes to be independently and identically distributed random variables $\in[0,2\pi]$. 
    We perform $8$ sets of DNSs: the first $3$ runs [Run$1$-Run$3$], with $512^{3}$ collocation points, and the next $3$ runs [Run$4$-Run$6$], with $1024^{3}$ collocation points, use $d_i=0.05$; the last $2$ runs [Run$7$-Run$8$], with $512^3$ collocation points, use $d_i=0.1$. The parameters for these runs provided initially and obtained at the cascade-completion time $t_{c}$ are given in Table \ref{table:1}. 
    
    To characterize the statistical properties of 3D HMHD turbulence, we compute the energy $E_{a}(k,t)$ and dissipation $\epsilon_{a}(k,t)$ spectra, the structure functions $S^{a}_{p}(l)$,
    and the flatnesses $F_{4}^{a}$ defined below [cf. Refs.\cite{sahoo2011systematics,yadav2022statistical}]:
\begin{eqnarray}
E_{a}(k,t) &\equiv& \sum \limits_{k-\frac{1}{2}\leq k^{\prime}\leq k+\frac{1}{2}}|\tilde{\bm a}({\bf k}^{\prime},t)|^{2}\,;\;\;
\epsilon_{a}(k,t)\equiv\nu_{a} k^{2}E_{a}(k,t)\,;\nonumber \\
E_{a}(t) &\equiv& \sum_k E_{a}(k,t)\,;\;\;
\epsilon_{a}(t) \equiv \sum_k \epsilon_{a}(k,t)\,;\;\;
\eta_{d}^{a} \equiv \left(\frac{\nu_{a}^{2}}{\epsilon_{a}}\right)^{1/4};\nonumber \\
\delta a_{\parallel}\left({\bm x},l\right) &\equiv& \left[{\bm a}({\bm x}+{\bm l},t)-{\bm a}({\bm x},t)\right]\cdot \frac{{\bm l}}{l}\,;\nonumber\\
 S^{a}_{p}(l)&\equiv& \langle |\delta a_{\parallel} (x,l)|^{p} \rangle\,;\;\; F_{4}^{a}\left(l\right)\equiv \frac{S_{4}^{a}\left(l\right)}{|S_{2}^{a}\left(l\right)|^{2}}\,;
 \label{eq:statproperties}
\end{eqnarray}
here, ${\bm a}$ is ${\bm {u}}$, for the velocity field, or ${\bm{b}}$, for the magnetic field, and $\nu_{u}(= \nu)$, the kinematic viscosity, and $\nu_{b}(= \eta)$, the magnetic diffusivity; $\eta_{d}^{u(b)}$ is the kinetic (magnetic) dissipation scale length; tildes denote spatial Fourier transforms, and 
$k^{\prime}$ is the modulus of the wave vector ${\bm k}^{\prime}$. We compute the time evolution of the energies $E_{a}(t)$ and the dissipation rates $\epsilon_{a}(t)$, and the $k$ dependence of the spectra $E_{a}(k,t)$ and the Alfv\'en ratio $E_{b}(k,t)/E_{v}(k,t)$
at the cascade-completion time $t_c$ (see below). Furthermore, we explore the 
dependence of $S^{a}_{p}$, $F_{4}^{a}$, and the probability distribution functions (PDFs) of the longitudinal-field increments $\delta a_{\parallel}$ on the length scale $l$.

The ideal 3D HMHD equations~\eqref{eq:vel}-\eqref{eq:divmag} have the following three quadratic invariants [see, e.g., Ref.~\cite{stawarz2015small}]:
\begin{eqnarray}
E_T &\equiv& E_{u}+ E_{b}\;\; [{\rm{total\; energy}}]\,;\nonumber \\
H_M &\equiv& \frac{1}{2}\langle {\bm{A}}\cdot{\bm{b}}\rangle\;\; [{\rm{magnetic\; helicity}}]\,;\nonumber \\
H_G &\equiv& H_M+2d_i H_C+d_{i}^{2} H_V\;\; [{\rm{generalized\; helicity}}]
 \label{eq:invariants}
\end{eqnarray}
here,  $H_{C}= \frac{1}{2}\langle \bm{u}\cdot\bm{b}\rangle $ is the cross helicity, $H_{V}= \langle \bm{u}\cdot\bm{\omega}\rangle $ the kinetic helicity,
and $\langle \cdot \rangle$ denotes the volume average. In ideal 3D MHD, the three quadratic invariants
are $E_T$, $H_M$, and $H_C$.


We recall that, if we linearize the incompressible 3D HMHD equations~\eqref{eq:vel}-\eqref{eq:divmag}, we find  that there are two wave modes, the right-circularly polarized (RCP) whistler-wave mode and the left-circularly polarized (LCP) ion-cyclotron wave mode [see, e.g., Refs.~\cite{krishan2004magnetic,stawarz2015small}]; and the $k$-dependent ratio of the magnetic energy to the fluid kinetic energy is  
\begin{equation}
\frac{E_{b} (k)}{E_{u} (k)}=\frac{k^{2}d_{i}^{2}}{4}\left(\pm 1 + \sqrt{1+\frac{4}{k^{2}d_{i}^{2}}}\right)^{2}\,,
\label{eq:alfvenratio}
\end{equation}
where the $+$ and $-$ signs are used, respectively, for the whistler and ion-cyclotron wave modes, whence we get, for
$kd_{i} \gg 1$,
\begin{eqnarray}
    E_{b}/E_{u} &\sim& k^{2}d_{i}^{2}\;\; [{\rm{whistler\; mode}}]\,;\label{eq:whis} \\
    E_{b}/E_{u} &\sim& \left(k^{2}d_{i}^{2}\right)^{-1}\;\; [{\rm{ion-cyclotron \;mode}}]\,.\label{eq:ionc}
\end{eqnarray}
Furthermore, the magnetic polarization is 
\begin{equation}
{\mathcal{P}}_m(k)\equiv \sigma_{m}(k) \sigma_{c}(k)\,,
\label{eq:magpol}
\end{equation}
where the $k$-dependent magnetic and cross helicities [averaged over shells of radius $k$ in Fourier space] are, respectively:
\begin{gather}
     \label{eq:maghelicity}
		\sigma_m(k)=\frac{\tilde{\bm{A}}\cdot\tilde{\bm{b}}^*+\tilde{\bm{A}}^*\cdot\tilde{\bm{b}}}{2|\tilde{\bm{A}}||\tilde{\bm{b}}|}\,;\\
    \label{eq:crosshelicity}
		\sigma_c(k)=\frac{\tilde{\bm{u}}\cdot\tilde{\bm{b}}^*+\tilde{\bm{u}}^*\cdot\tilde{\bm{b}}}{2|\tilde{\bm{u}}||\tilde{\bm{b}}|}\,;
	\end{gather}
    here, asterisks denote complex conjugation. 
    The magnetic polarization ${\mathcal{P}}_{m}\in~[-1,1]$. It is negative for whistler waves and positive for ion-cyclotron waves, which we denote, henceforth, by the subscripts $R$ and $L$, because they are associated, respectively, with RCP and LCP fluctuations [see, e.g., Ref.~\cite{sahraoui2020magnetohydrodynamic}], whose energy spectra and their sum are:
    \begin{gather}
    \label{eq:rightleftfluct}
    E_{R,L}(k,t)\equiv[E_u(k,t)+E_b(k,t)]_{R,L};\\
	\label{eq:totalrightandleftfluct} 
	E_{T}(k,t)\equiv[E_R(k,t)+E_L(k,t)].
    \end{gather}
	\section{Results}
    \label{sec:Results}
    We now present the results from our DNSs [Run1-Run8 in Table~\ref{table:1}].
     In Section~\ref{subsec:tevol}, we describe the time evolution of the energy dissipation rates and glean insights from iso-surface plots of the moduli of the vorticity and the current density. Section~\ref{subsec:spectra} presents fluid [$E_{u}(k)$] and magnetic-field [$E_{u}(k)$] spectra  as functions of the wavenumber $k$ and also spectra that are conditioned on their $L$ or $R$ polarization. In Section~\ref{subsec:inter} we characterize the intermittency of 3D HMHD turbulence using order-$p$ structure functions, flatnesses, and the probability distribution functions (PDFs) of velocity- and magnetic-fields increments.
    
	\begin{table*}[htbp]
		\setlength{\tabcolsep}{10pt}
		\begin{tabular}{c c c c c c c c c c c c}
			\hline
			\hline
			Runs & $N$ & $\nu$ & $Pr_m$ & $d_i$ & $\delta t$ & $u_{rms}$ & $\lambda$ & $Re_{\lambda}$  & $t_c$ & $k_{max}\eta^u_d$ & $k_{max}\eta^b_d$ \\			\hline
			Run1 & 512 & $5\times10^{-4}$ & 0.1 & 0.05 & $10^{-4}$ & 0.36 & 0.43 & 253.20 & 6.92 & 1.90 & 6.84\\
			
			Run2 & 512 & $10^{-3}$ & 1 & 0.05 & $10^{-4}$ & 0.36 & 0.32 & 116.63 & 7.05 & 2.56 & 2.23\\
			
			Run3 & 512 & $5\times10^{-3}$ & 10 & 0.05 & $10^{-4}$ & 0.35 & 0.36 & 37.10 & 5.69 & 7.37 & 1.53\\
			
			Run4 & 1024 & $5\times10^{-4}$ & 0.1 & 0.05 & $5\times10^{-5}$ & 0.32 & 0.39 & 211.38 & 6.93 & 3.83 & 13.72\\	
			
			Run5 & 1024 & $5\times10^{-4}$ & 1 & 0.05 & $10^{-5}$ & $0.31$ & $0.21$ & $129.83$ & $8.23$ & $3.16$ & $2.72$\\
			
			Run6 & 1024 & $5\times10^{-3}$ & 10 & 0.05 & $10^{-5}$ & $0.30$ & $0.34$ & $28.30$ & $7.21$ & $15.40$ & $3.16$\\		
			
			Run7 & 512 & $5\times10^{-4}$ & 0.1 & 0.1 & $5\times10^{-5}$ & $0.38$ & $0.45$ & $292.51$ & $6.95$ & $1.90$ & $6.85$\\
			
			

			Run8 & 512 & $5\times10^{-3}$ & 10 & 0.1 & $5\times10^{-5}$ & $0.39$ & $0.35$ & $43.84$ & $5.73$ & $7.32$ & $1.47$\\											
			\hline
			\hline
		\end{tabular}
		
		\caption{\small Parameters for our DNS runs [Run$1$-Run$8$] for the initial $(t=0)$ spectra [see Eq.~\eqref{eq:initialspectra} and Fig.~\ref{fig:1}. Given our $2/3$-dealiasing method~\cite{canuto2012spectral,sahoo2011systematics}, the maximal wave numbers are $k_{max}=171$ and $342$, for runs with $N^{3}=512^{3}$ and $1024^{3}$ collocation points, respectively. Parameters: $\nu$, kinematic viscosity; $\eta$,  magnetic diffusivity; $Pr_{m}=\nu/\eta$, magnetic Prandtl number; $d_{i}$, ion-inertial length. The following quantities are computed at $t_c$, the cascade completion time (see text): $u_{rms}$, root-mean-square velocity; $\lambda$, Taylor-microscale; $Re_{\lambda}$, Taylor-microscale Reynolds number; $\eta_{d}^{u}$ and $\eta_{d}^{b}$ are, respectively, the kinetic and magnetic dissipation length scales (see text).}
		\label{table:1}
	\end{table*}
    
	\subsection{Temporal Evolution}
    \label{subsec:tevol}
\begin{figure*}[t]
\centering 
\begin{tabular}{c c c}
\text{(a)} & \text{(b)} & \text{(c)}  \\
\includegraphics [scale=0.22]{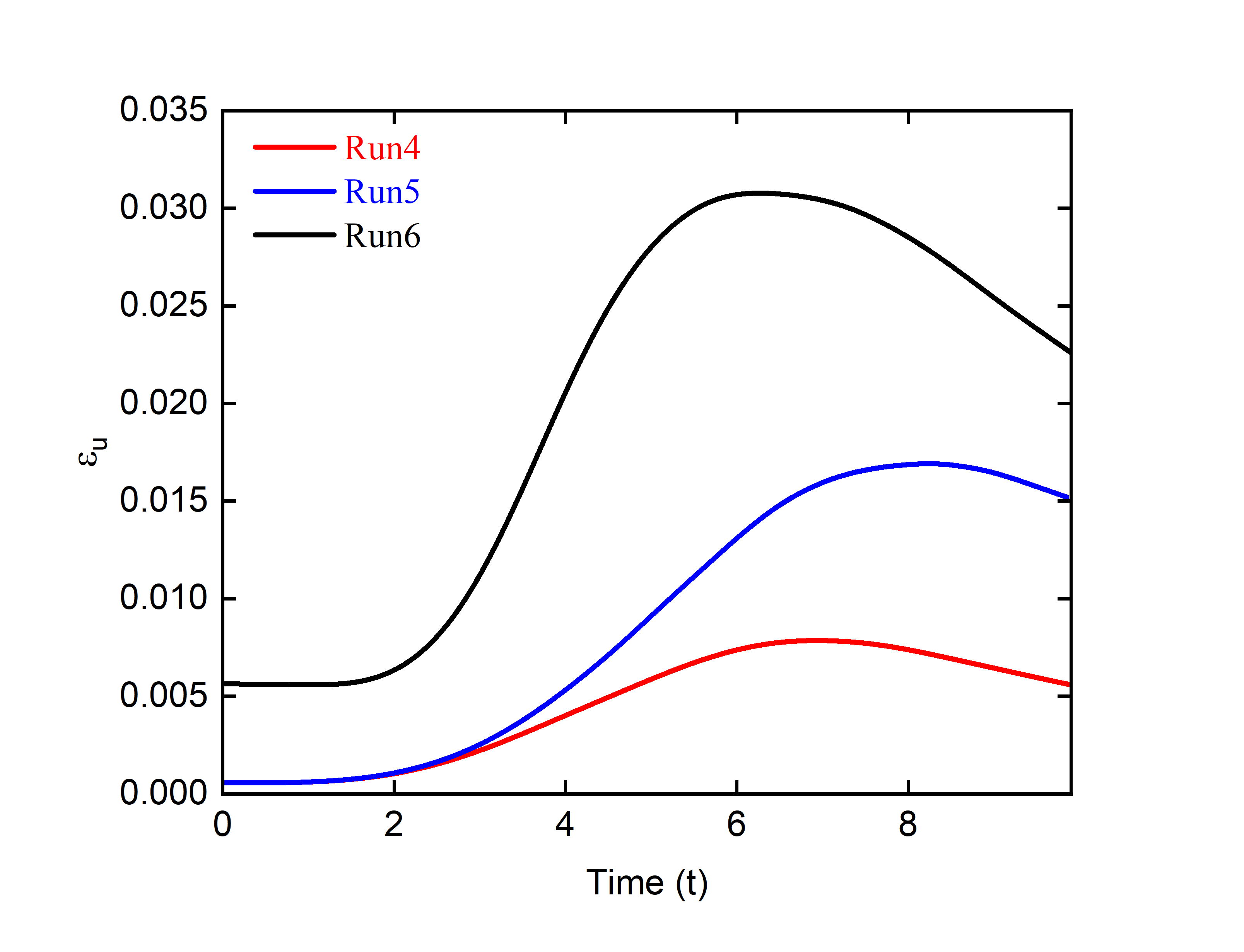} & 
\includegraphics [scale=0.22]{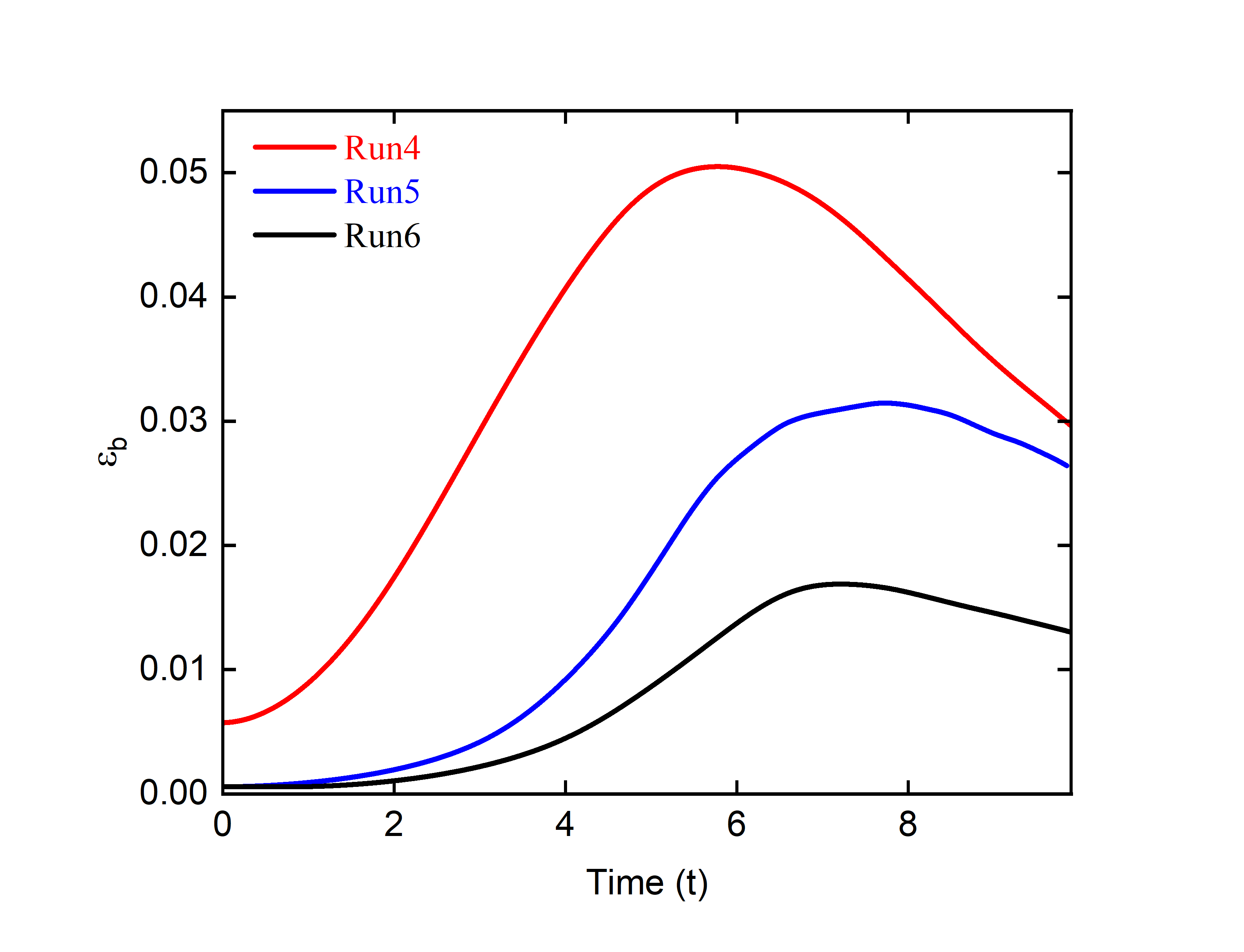} &
\includegraphics [scale=0.22]{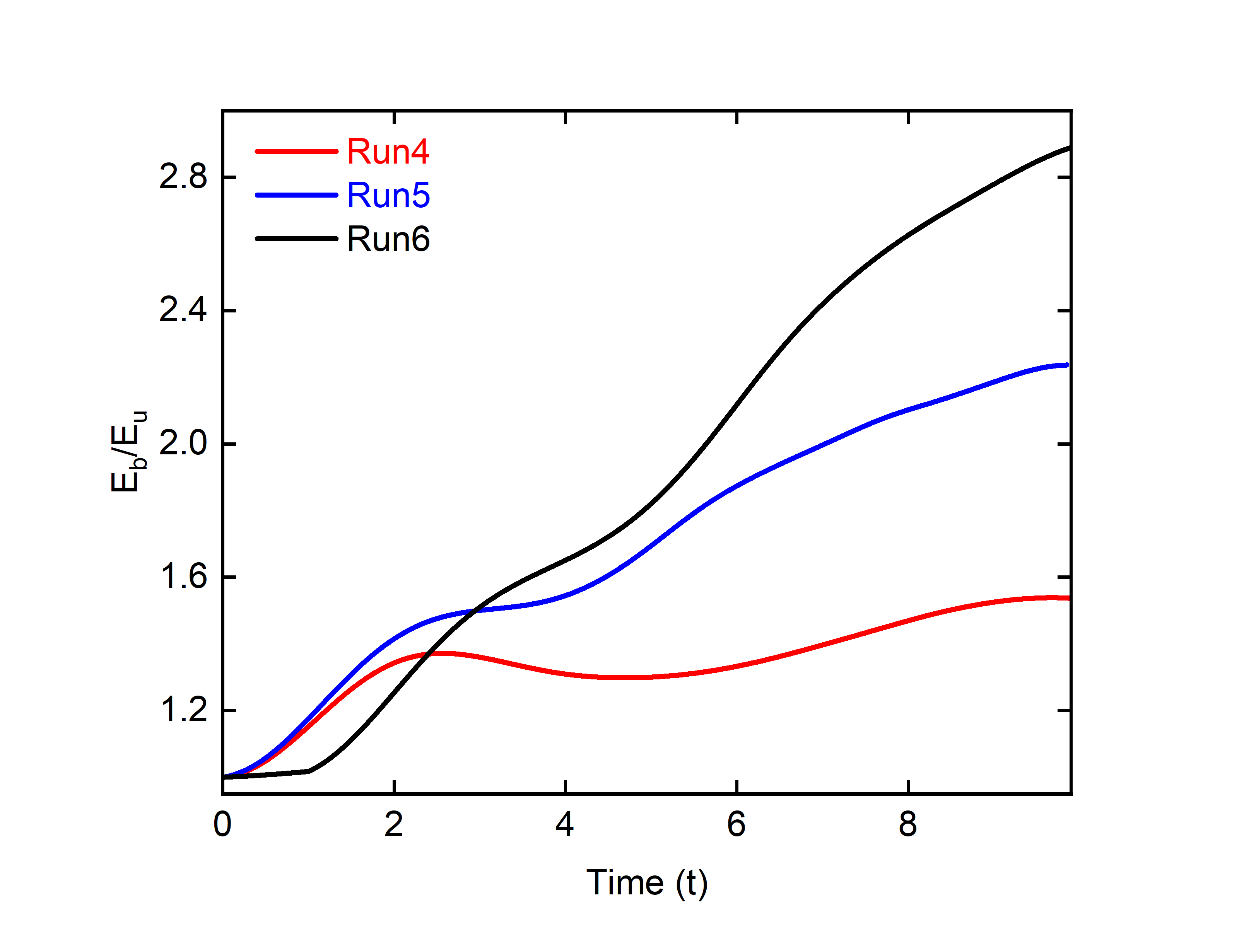}
\end{tabular}
\vskip -0.1cm 
	\caption{
    Plots  {\it{versus}} time $t$ of the (a) fluid-energy dissipation rates per unit mass $\epsilon_\text{u}$, (b) magnetic-energy dissipation rates per unit mass $\epsilon_\text{b}$, and (c) the total Alfv\'en ratios $E_b/E_u$ for Run4 (red curves), Run5 (blue curves), and Run6 (black curves) with $Pr_{m}=0.1$, $1$, and $10$, respectively.}
    \label{fig:tempevol}
\end{figure*}

In Fig.~\ref{fig:tempevol}, we characterize the temporal evolution of the fluid and magnetic energy-dissipation rates $\epsilon_{u}$ and $\epsilon_{b}$, respectively,  and the total Alfv\'en ratio, for Run4, Run5, and Run6, with 
 $Pr_{m}=0.1$, $1$, and $10$, respectively. Figures~\ref{fig:tempevol} (a) and (b) show  plots versus the time $t$ of $\epsilon_{u}$ and $\epsilon_{b}$. The curves in these plots increase initially, reach a peak, and then decay gradually; within a few time steps from the beginning of our simulations, they satisfy:  
\begin{eqnarray}
\epsilon^{Pr_{m}=0.1}_{u} &>& \epsilon^{Pr_{m}=1.0}_{u} > \epsilon^{Pr_{m}=10.0}_{u}\,;\nonumber \\
\epsilon^{Pr_{m}=0.1}_{b} &<& \epsilon^{Pr_{m}=1.0}_{b} < \epsilon^{Pr_{m}=10.0}_{b}\,;
\label{eq:epstrend}
\end{eqnarray}
these trends are also found in decaying MHD turbulence~\cite{sahoo2011systematics}. The peaks in the plots 
of the energy-dissipation rates yield an estimate for the \textit{cascade-completion time} $t_{c}$. In Fig.~\ref{fig:tempevol} (c), we show that the total Alfv\'en ratio $E_{b}/E_{u}$ increases with $t$ over the durations of Run4, Run5, and Run6; furthermore, this ratio increases with $Pr_{m}$. 

\begin{figure*}[t]
\centering 
\begin{tabular}{c c c}
\text{(a)} & \text{(b)} & \text{(c)}  \\
\includegraphics [scale=0.17]{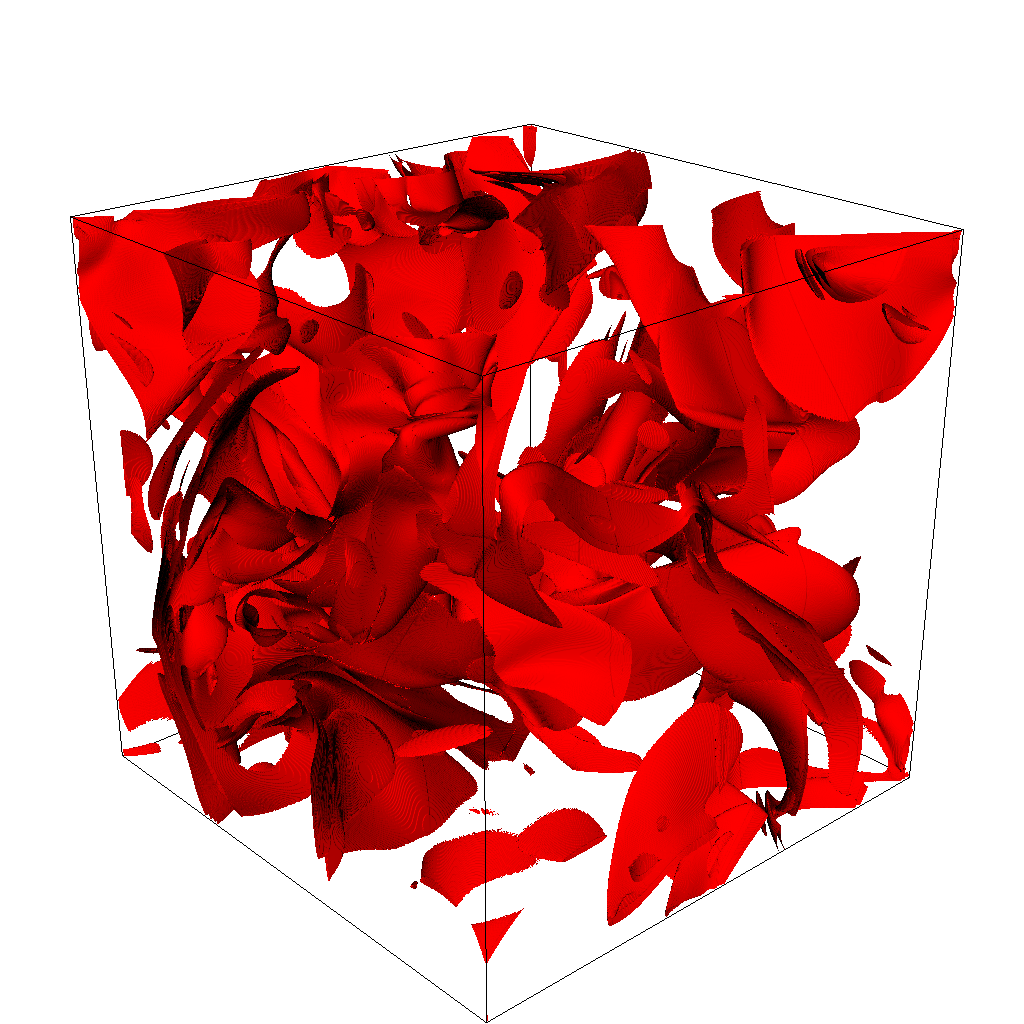} & 
\includegraphics [scale=0.17]{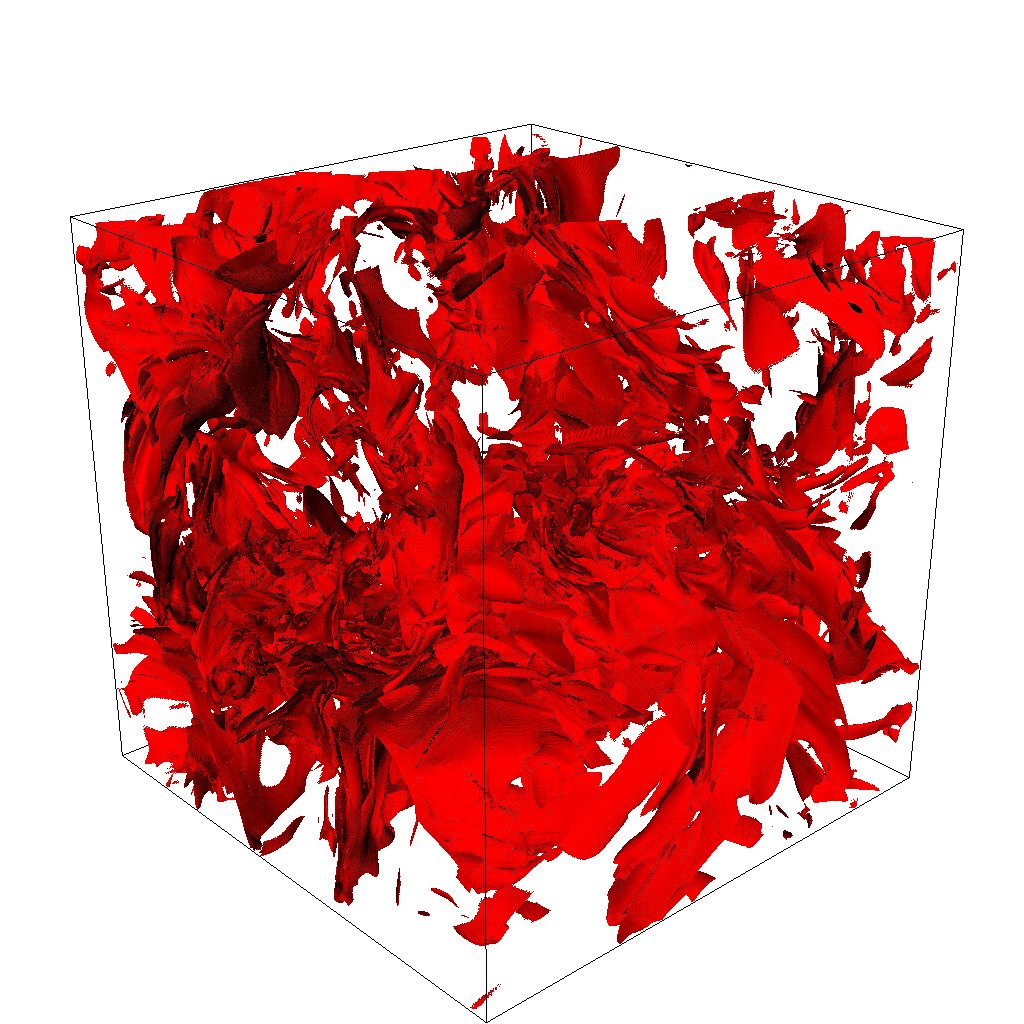} &
\includegraphics [scale=0.17]{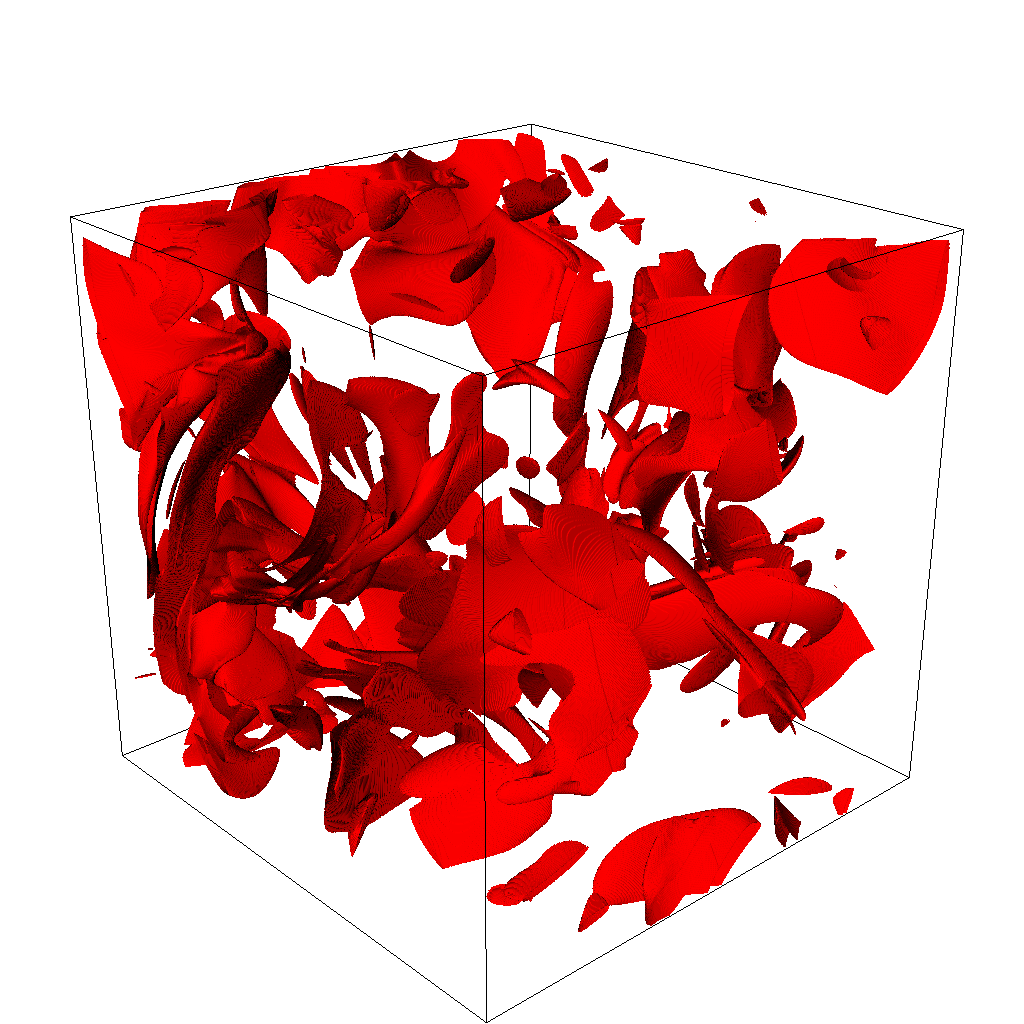}
\end{tabular}
%
\begin{tabular}{c c c}
\text{(d)} & \text{(e)} & \text{(f)}  \\
\includegraphics [scale=0.17]{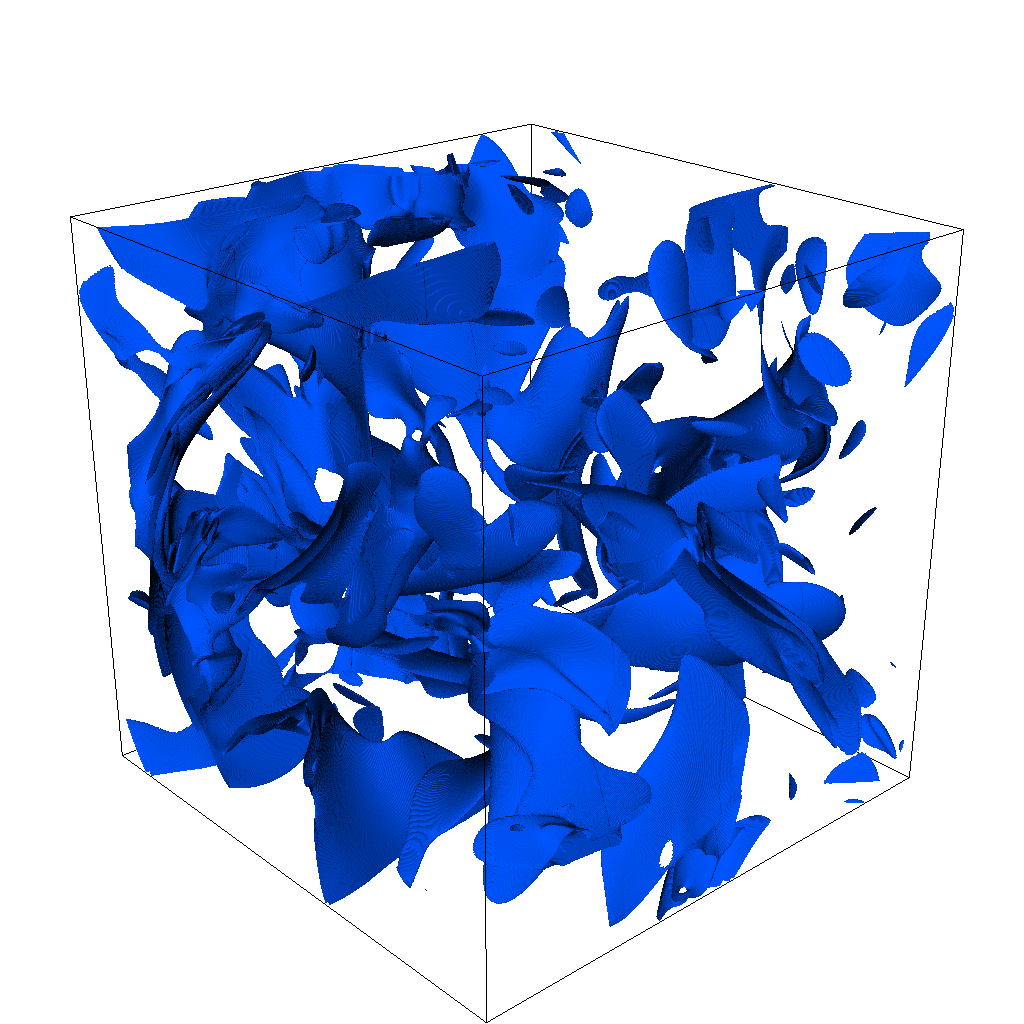} &
\includegraphics [scale=0.17]{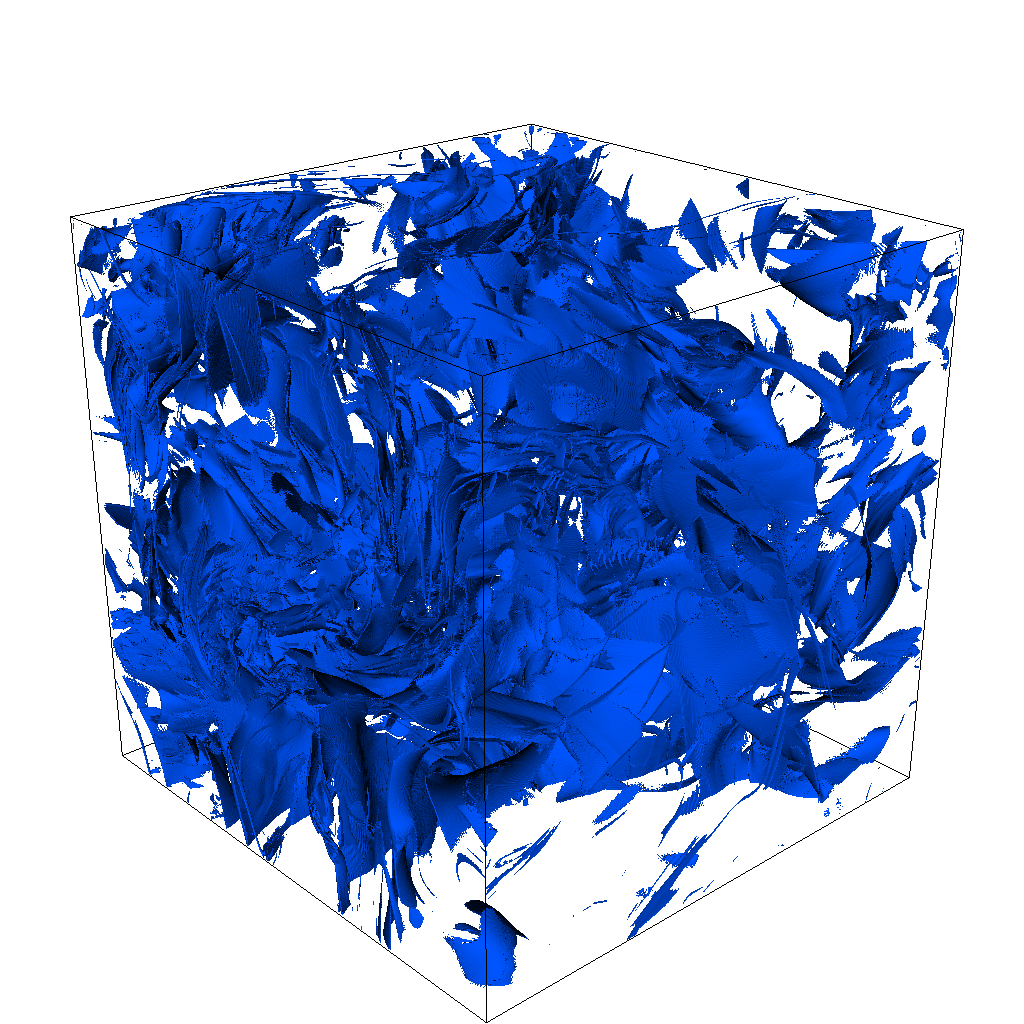} & 
\includegraphics [scale=0.17]{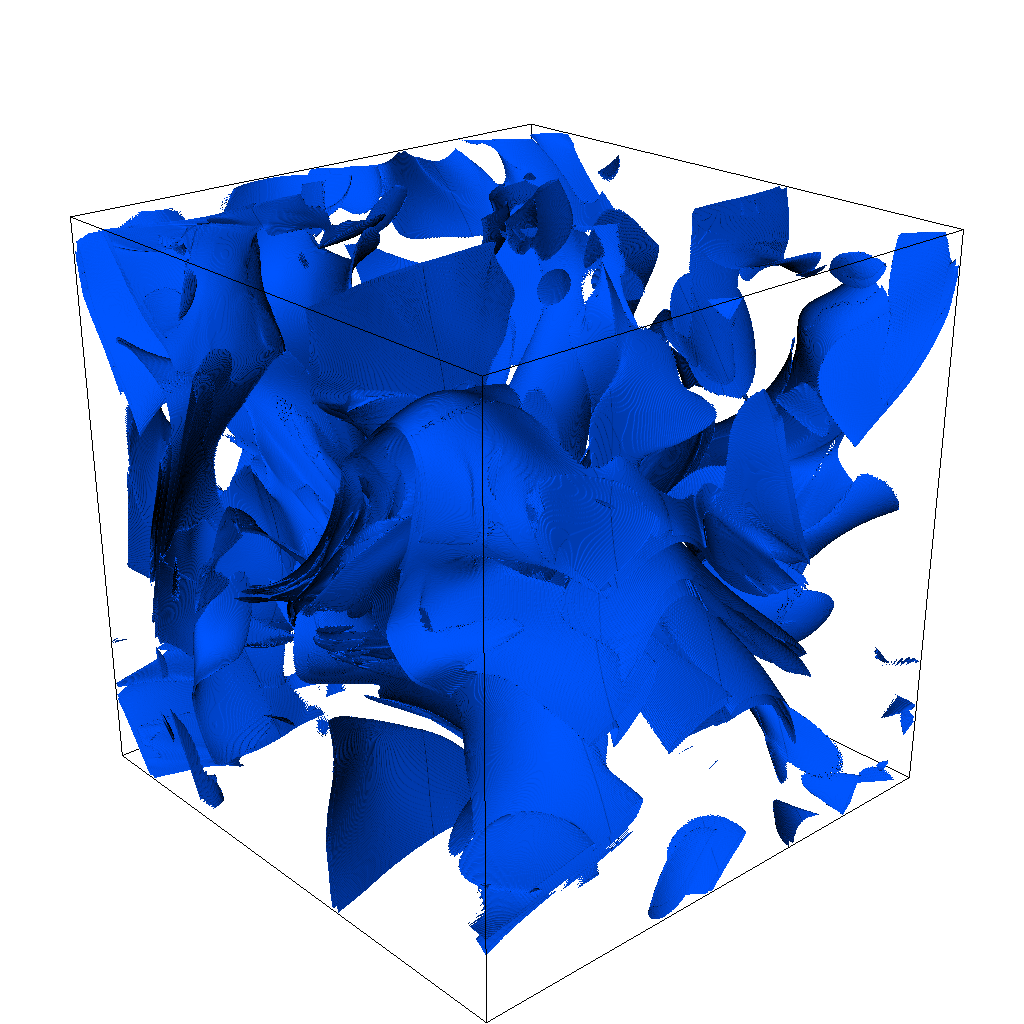}
\end{tabular}
\vskip -.1cm

\caption{Isosurfaces of the moduli of the vorticity $|\bm \omega|$ (in red) and the current density $|\bm J|$ (in blue) for: (a) and (d) Run4 [$Pr_{m} =0.1$]; (b) and (e) Run5 [$Pr_{m} =1$]; and (c) and (f) Run6 [$Pr_{m} =10$]. }

\label{fig:omegajisosurfacesrun456}
\end{figure*}

For decaying turbulence, in general, and 3D HMHD turbulence, in particular, it is natural to present statistical properties at 
the cascade-completion time $t_c$. In Fig.~\ref{fig:omegajisosurfacesrun456} we display isosurfaces of the moduli of the vorticity $|\bm \omega|$ (in the top row in red) and the current density $|\bm J|$ (in the bottom row in blue) for Run4 [$Pr_{m} =0.1$], Run5 [$Pr_{m} =1$], and Run6 [$Pr_{m} =10$]. These isosurfaces show fine-scale structures at the ion-inertial scale $d_i$, for Run5 [$Pr_{m} =1$], which are suppressed in Run4 [$Pr_{m} =0.1$] and Run6 [$Pr_{m} =10$], because of the high magnetic diffusivity, in the former, and the large viscosity, in the latter. This suppression gives us a hint of the subtle interplay between the magnetic and kinetic dissipation length scales, $\eta_{d}^{b}$ and $\eta_{d}^{u}$, and the ion-inertial scale $d_i$. We quantify this competition between these three length scales by computing various energy spectra in the next Section~\ref{subsec:spectra}.



%
%
\begin{figure*}
	\centering
		\includegraphics[scale=0.35]{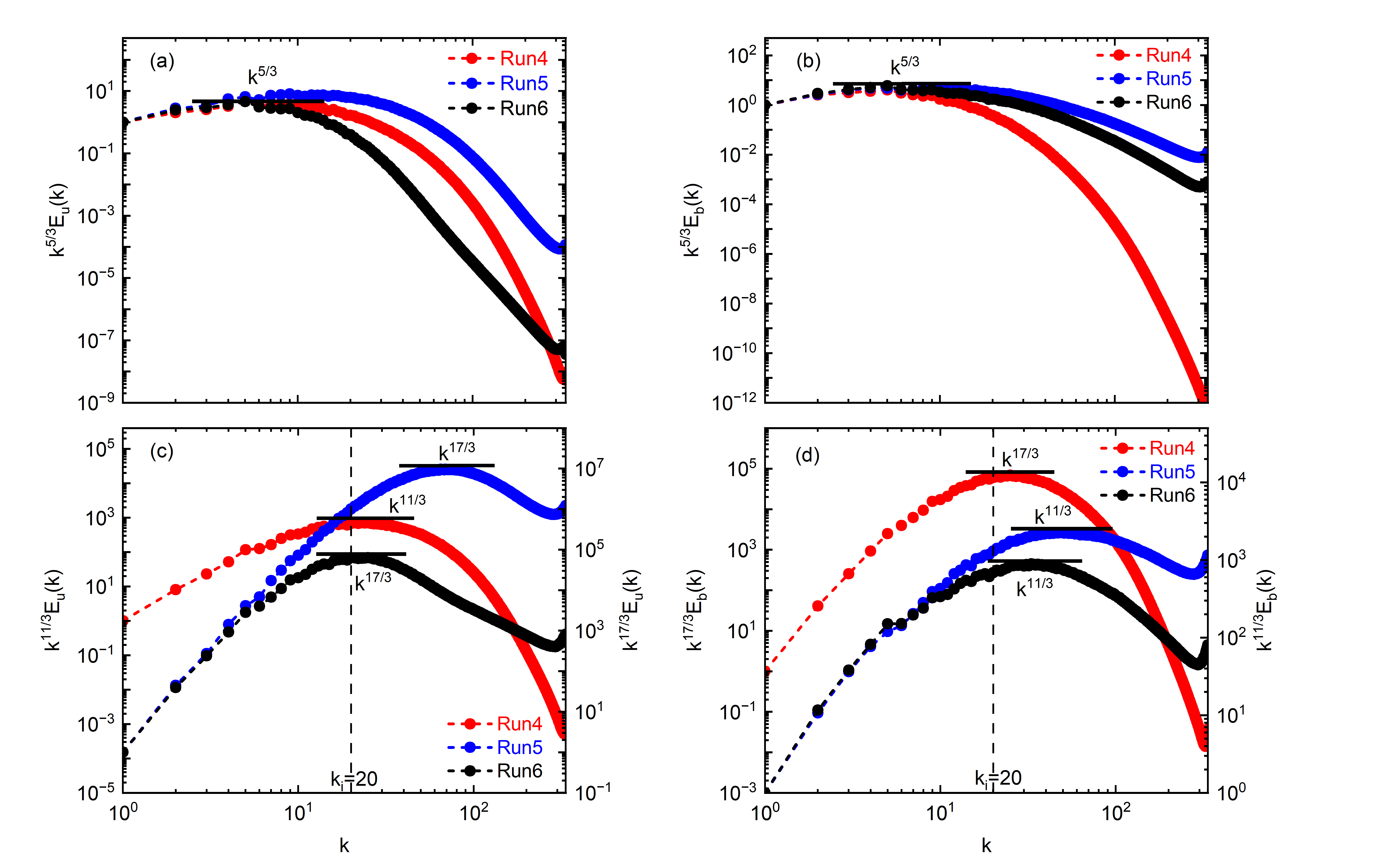}\hfill
	\centering
 \caption{\small Log-log (base 10) plots versus the wave number $k$ of compensated fluid- and magnetic-energy spectra: (a) $k^{5/3}E_u(k)$; (b) $k^{5/3}E_b(k)$; the compensated parts of the spectra are shown by  black-horizontal-line in the inertial regions.  (c) the compensated kinetic energy spectra with different power-laws $k^{\alpha}E_u(k)$ ($\alpha=11/3$ and $17/3$), and (d) the compensated magnetic energy spectra with different power-laws $k^{\alpha}E_b(k)$ ($\alpha=11/3$ and $17/3$) in the sub-inertial region, for Run4 (dash-red-curves), Run5 (dash-blue-curves), and Run6 (dash-black-curves); Run4, Run5 and Run6 have $Pr_{m}=0.1$, $1.0$, and $10.0$, respectively. }
	\label{fig:energyspectra}
\end{figure*}

\begin{figure*}[t]
\centering 
\begin{tabular}{c c c}
\text{(a)} & \text{(b)} & \text{(c)}  \\
\includegraphics [scale=0.22]{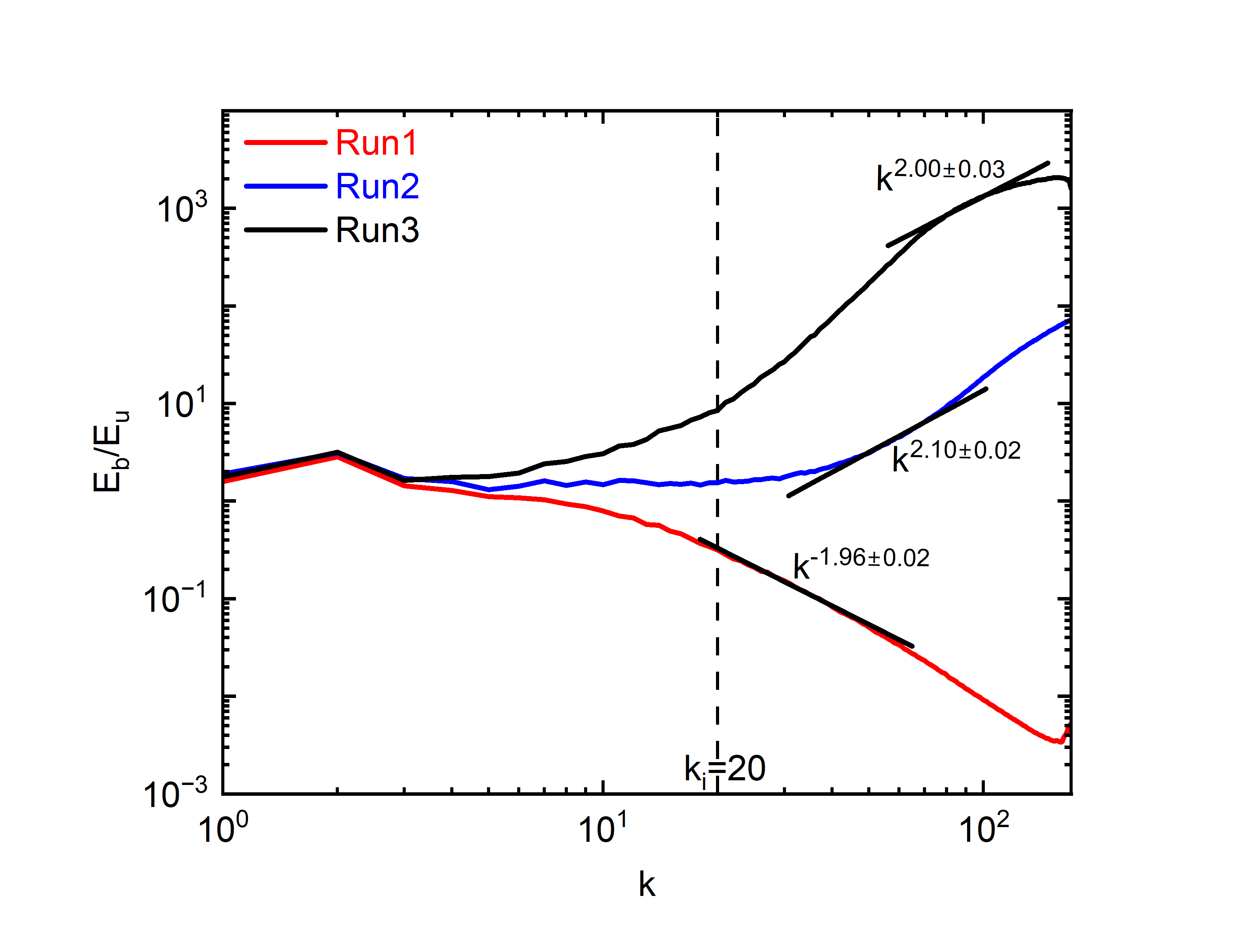} & 
\includegraphics [scale=0.22]{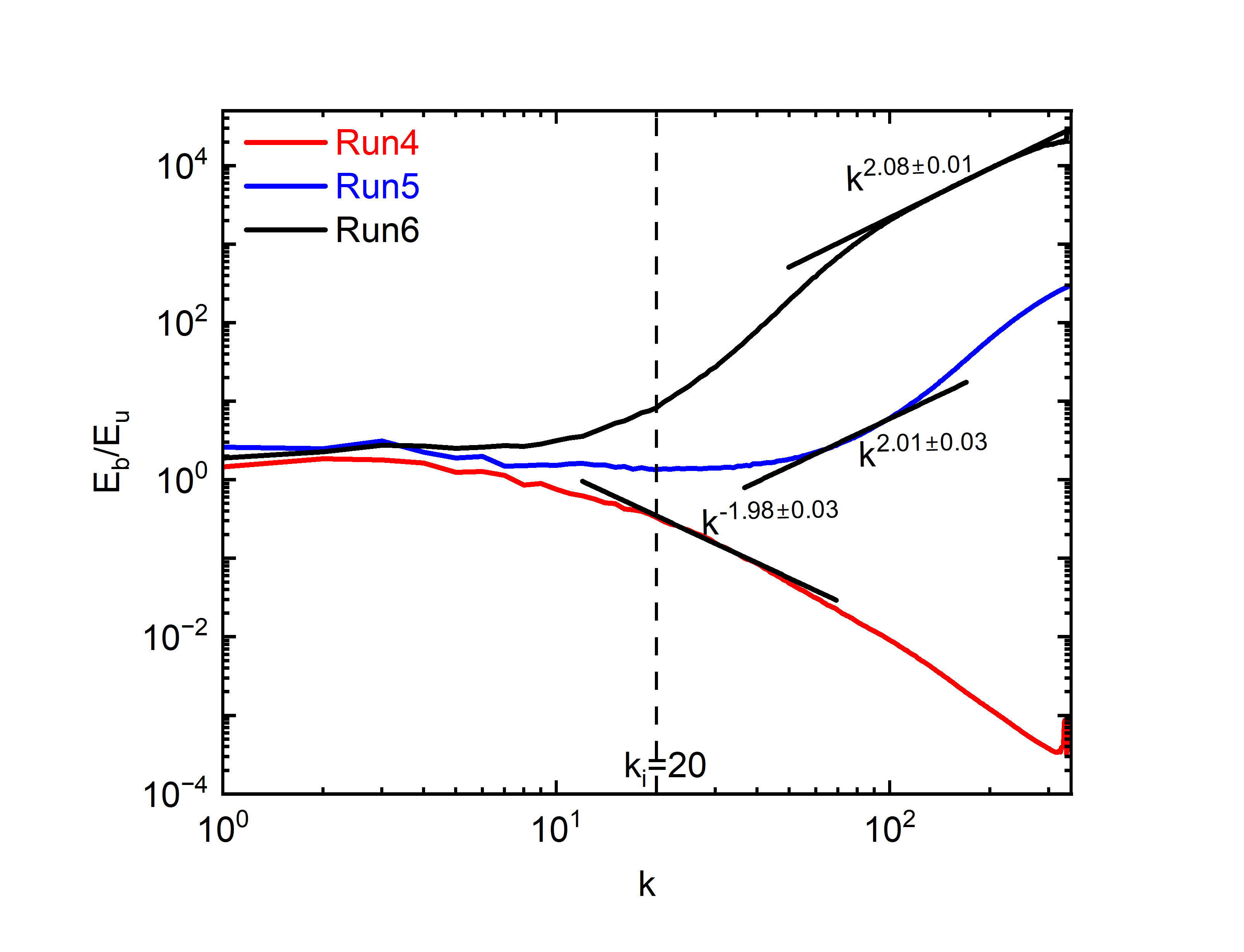} &
\includegraphics [scale=0.22]{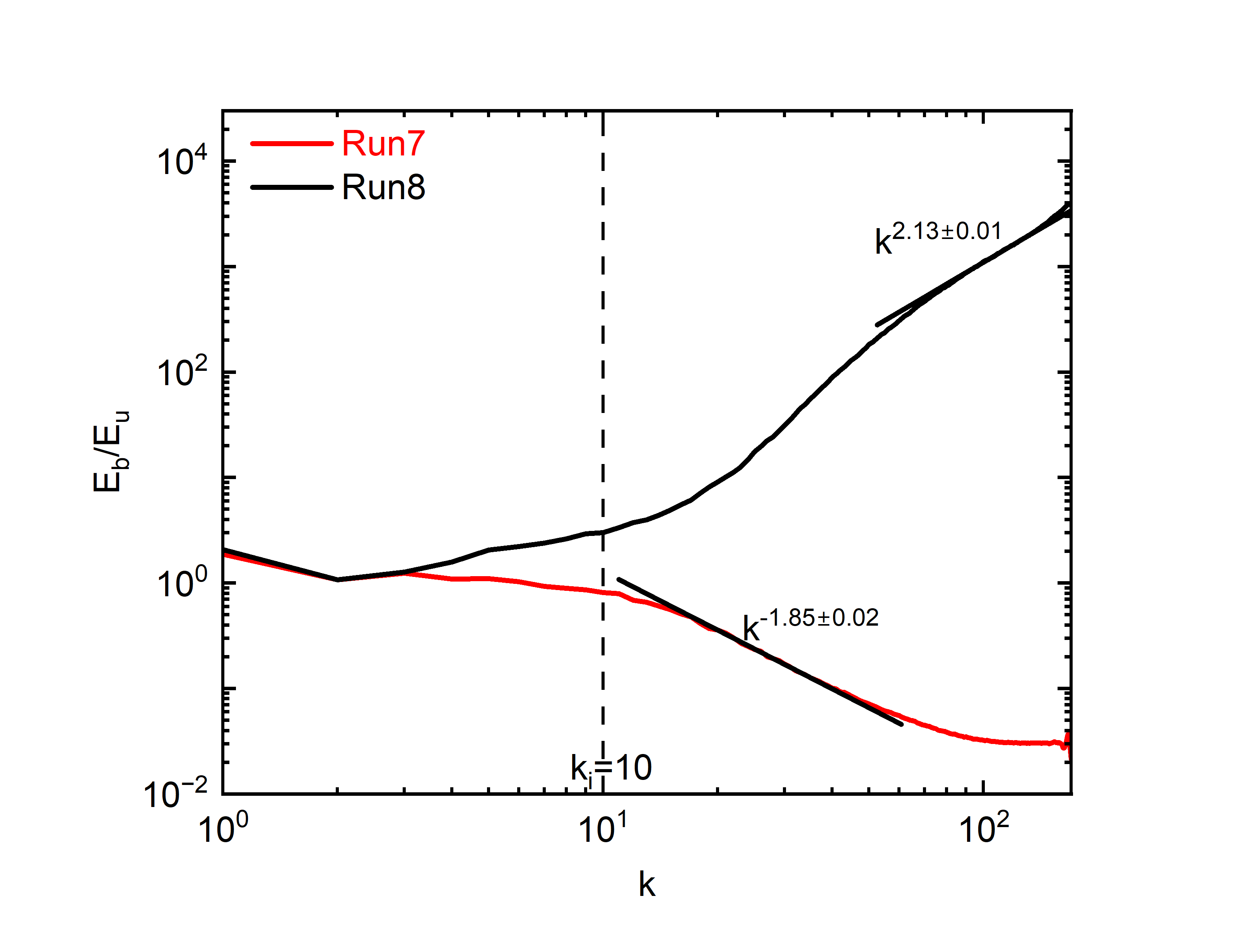}
\end{tabular}
\vskip -0.1cm 
	\caption{
    Log-log plots versus the wavenumber $k$ of the Alfv\'en ratios $E_b(k,t_c)/E_u(k,t_c)$ (a) for Run1 (solid-red-line), Run2 (solid-blue-line), and Run3 (solid-black-line), (b) for Run4 (solid-red-line), Run5 (solid-blue-line), and Run6 (solid-black-line), and (c) for Run7 (solid-red-line) and Run8 (solid-black-line), at the cascade-completion time; the black lines on the each curve of it in the sub-inertial indicate the region that is used for the computation of the slopes. In spite of the  limited spectral range, our DNSs display the interplay between $d_i$ and the two dissipation scales.}
    \label{fig:alfvenratiospectra}
\end{figure*} 
%
%
%
%
\begin{figure*}
	\centering
		\includegraphics[scale=0.25]{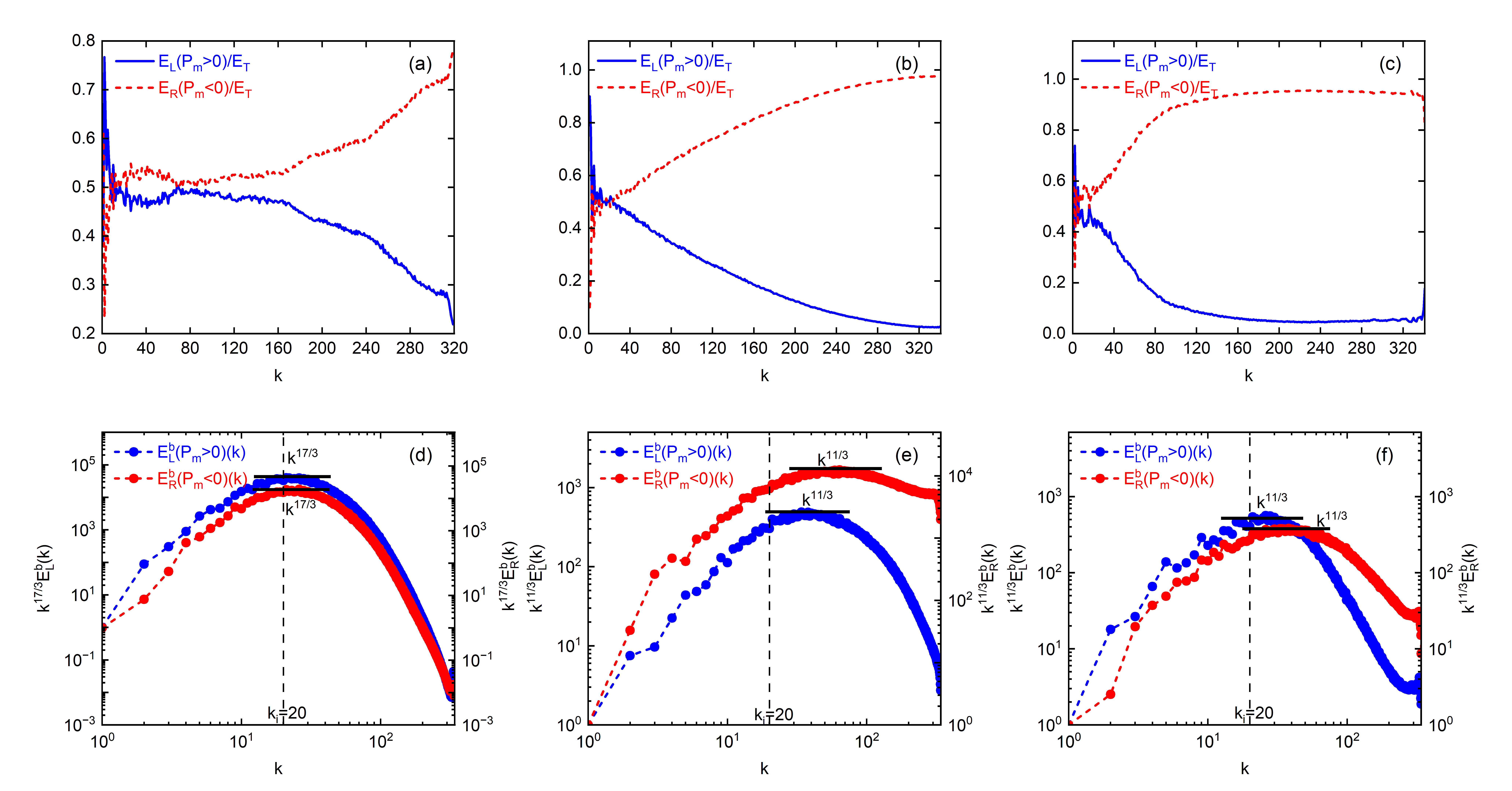}\hfill
	\centering
 \caption{\small Plots of the energy ratios $E_L/E_T$ (solid line) and $E_R/E_T$ (dashed line) for (a) Run4, (b) Run5, and (c) Run6. The compensated polarized magnetic energy spectra for $L$ fluctuations ${\mathcal{P}}_m(k)>0$ (blue curve) and $R$ fluctuations ${\mathcal{P}}_m(k)<0$ (red curve) for (d) Run4, (e) Run5, and (f) Run6.  }
	\label{fig:fluctuationrun456}
\end{figure*}
\begin{figure*}
	\centering
		\includegraphics[scale=0.35]{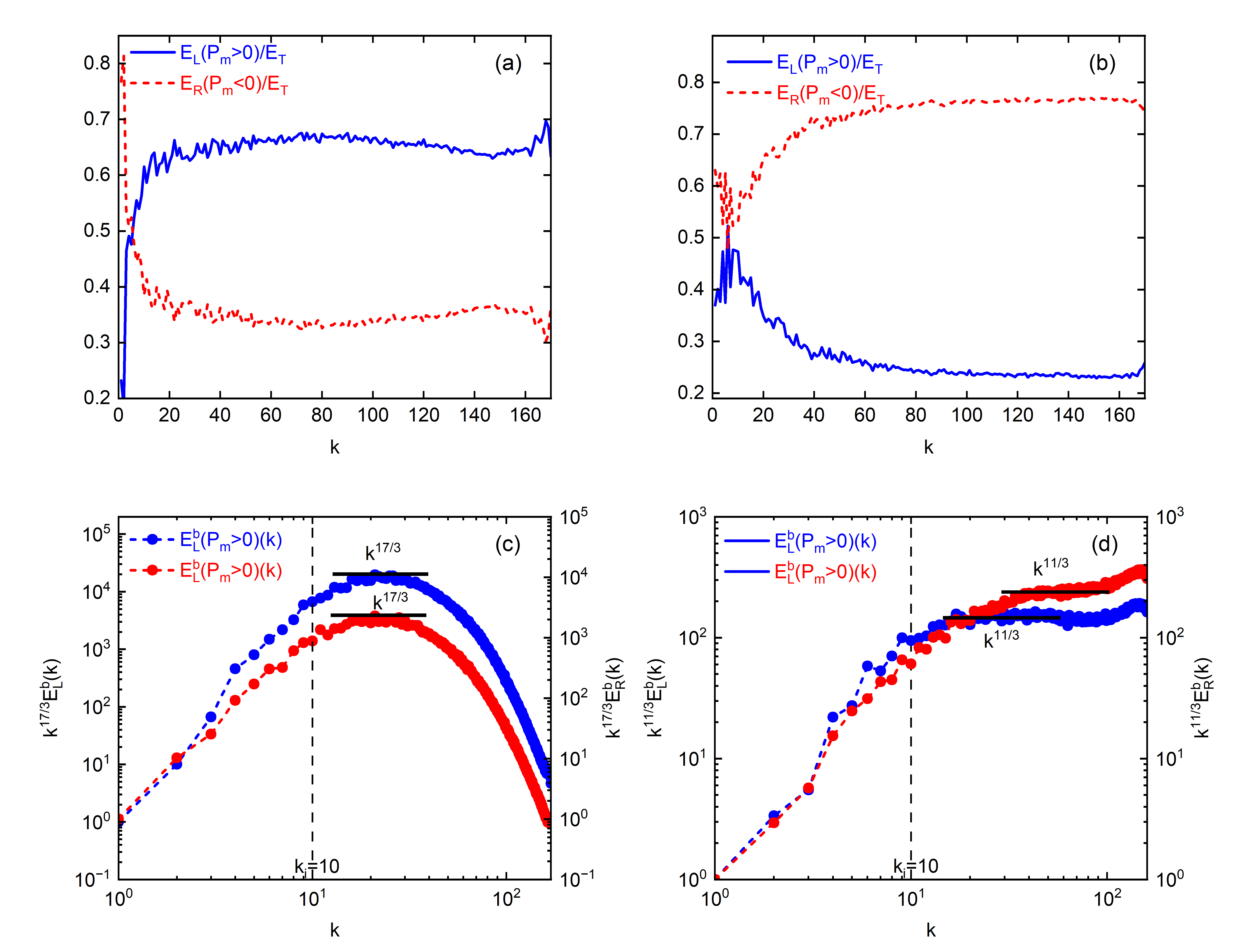}\hfill
	\centering
 \caption{\small Plots of the energy ratios $E_L/E_T$ (solid line) and $E_R/E_T$ (dashed line) for (a) Run7 and (b) Run8. The compensated polarized magnetic energy spectra for $L$ fluctuations  ${\mathcal{P}}_m(k)>0$  (blue curve) and $R$ fluctuations   ${\mathcal{P}}_m(k)<0$  (red curve) for (c) Run7 and (d) Run8. }
	\label{fig:fluctuationrun78}
\end{figure*}
\begin{figure*}[t]
\centering 
\begin{tabular}{c c c}
\text{(a)} & \text{(b)} & \text{(c)}  \\
\includegraphics [scale=0.22]{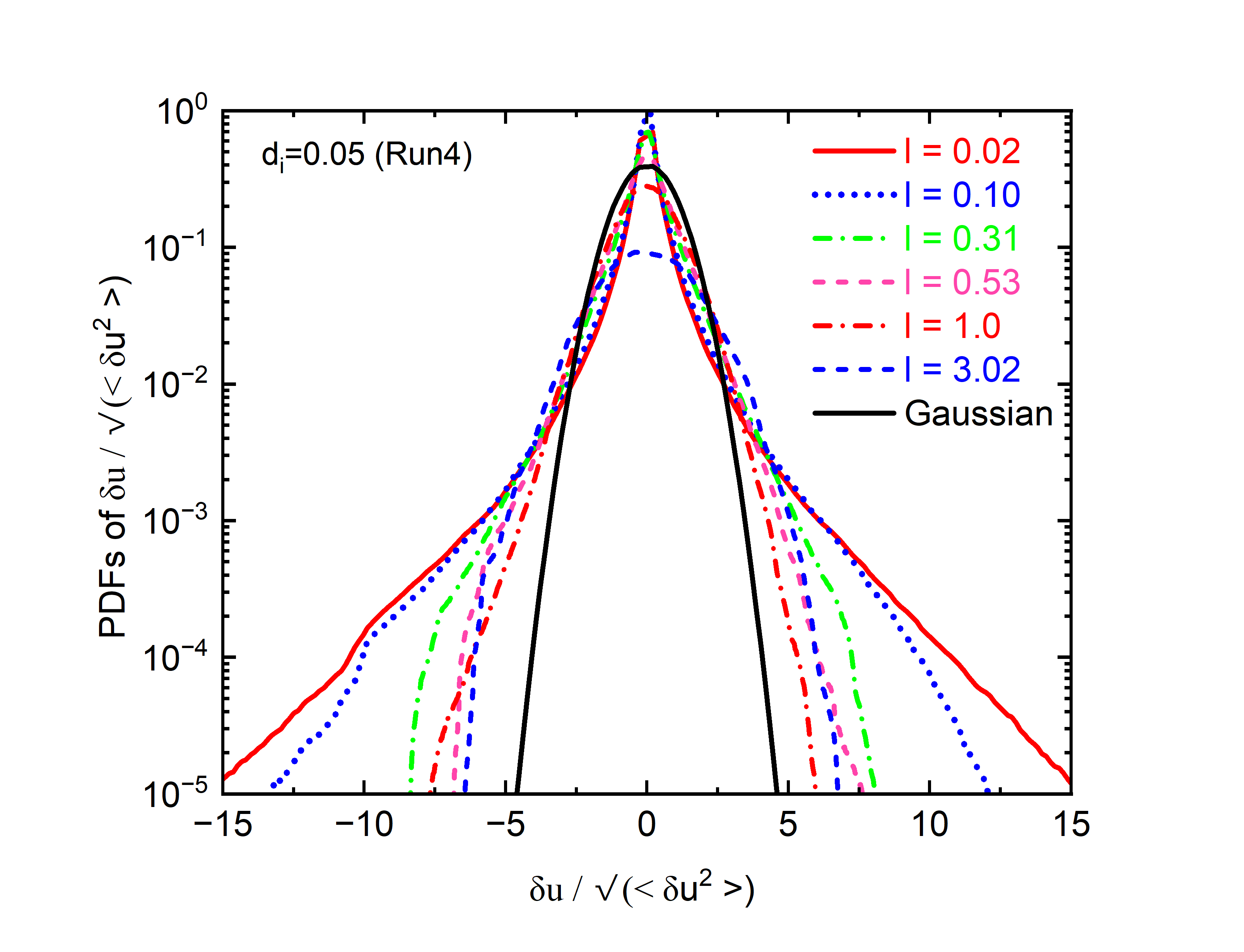} & 
\includegraphics [scale=0.22]{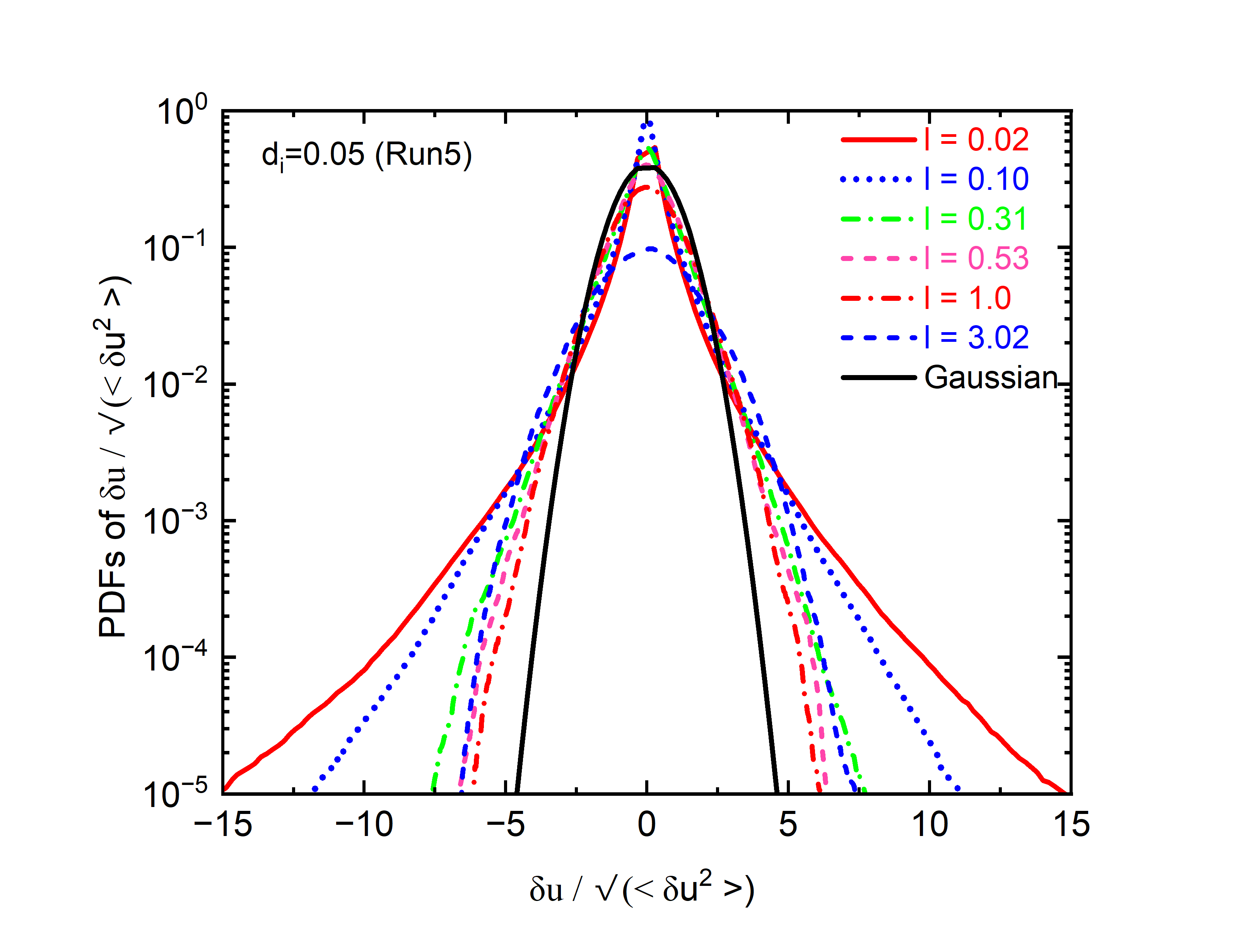} &
\includegraphics [scale=0.22]{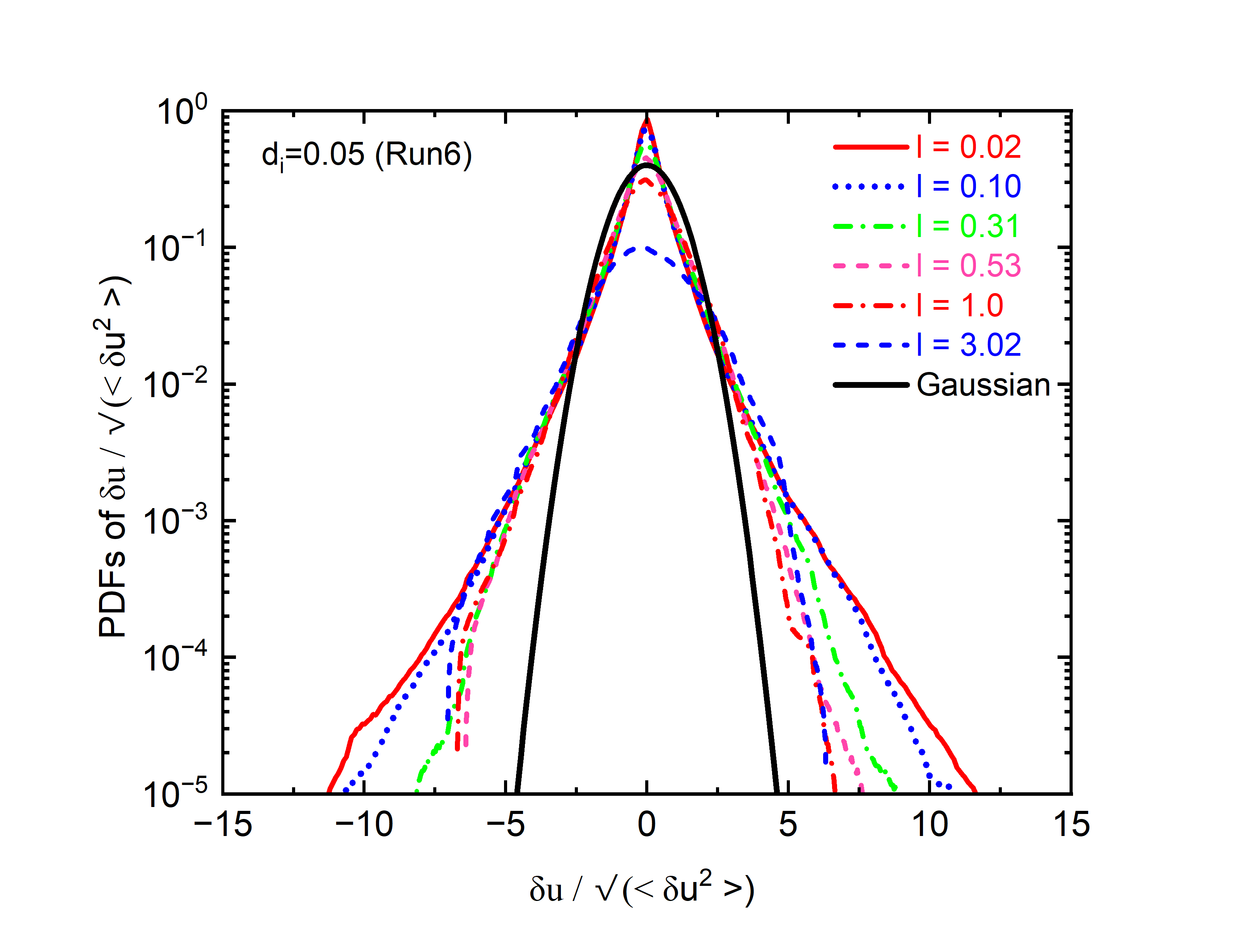}
\end{tabular}
%
\begin{tabular}{c c c}
\text{(d)} & \text{(e)} & \text{(f)}  \\
\includegraphics [scale=0.22]{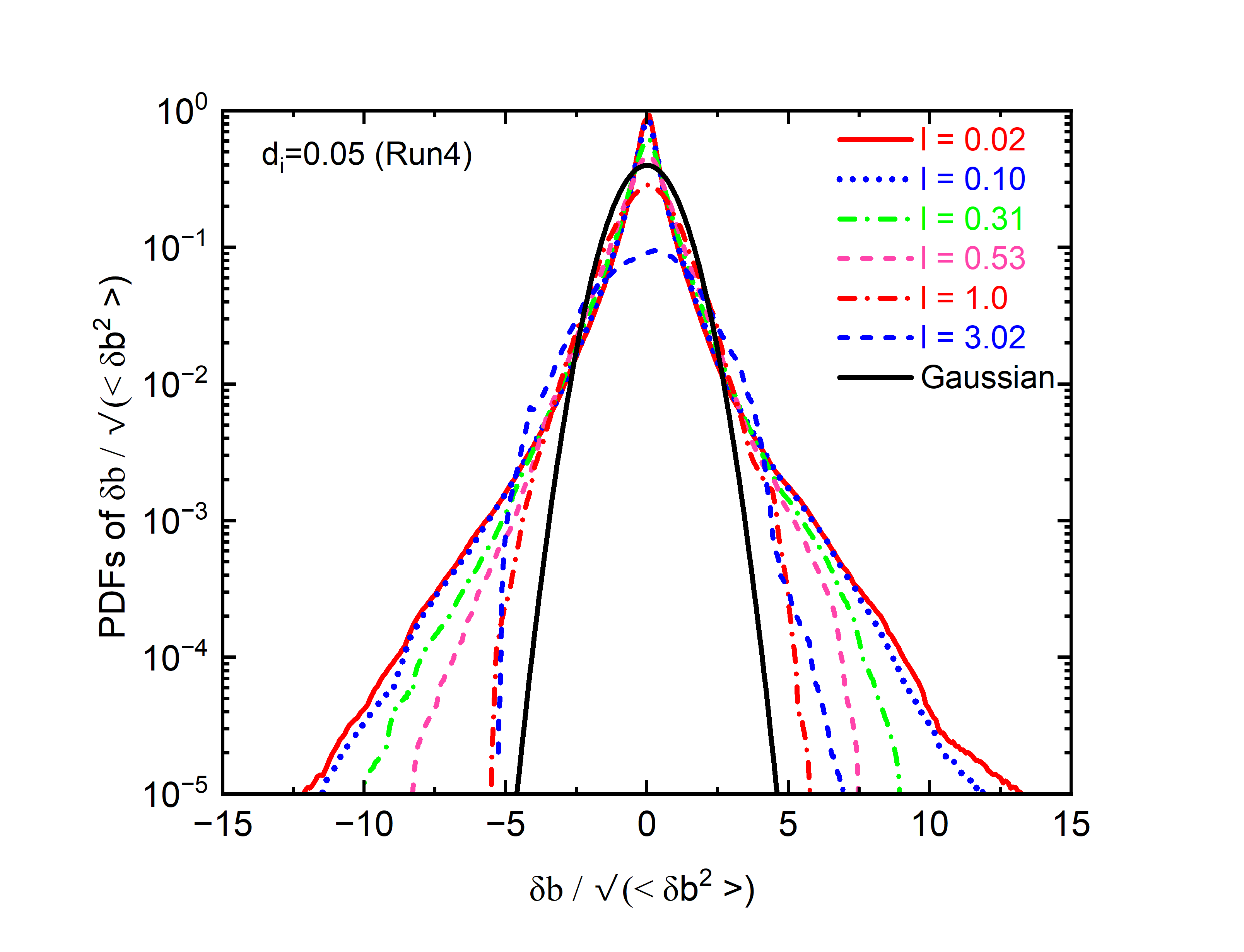} &
\includegraphics [scale=0.22]{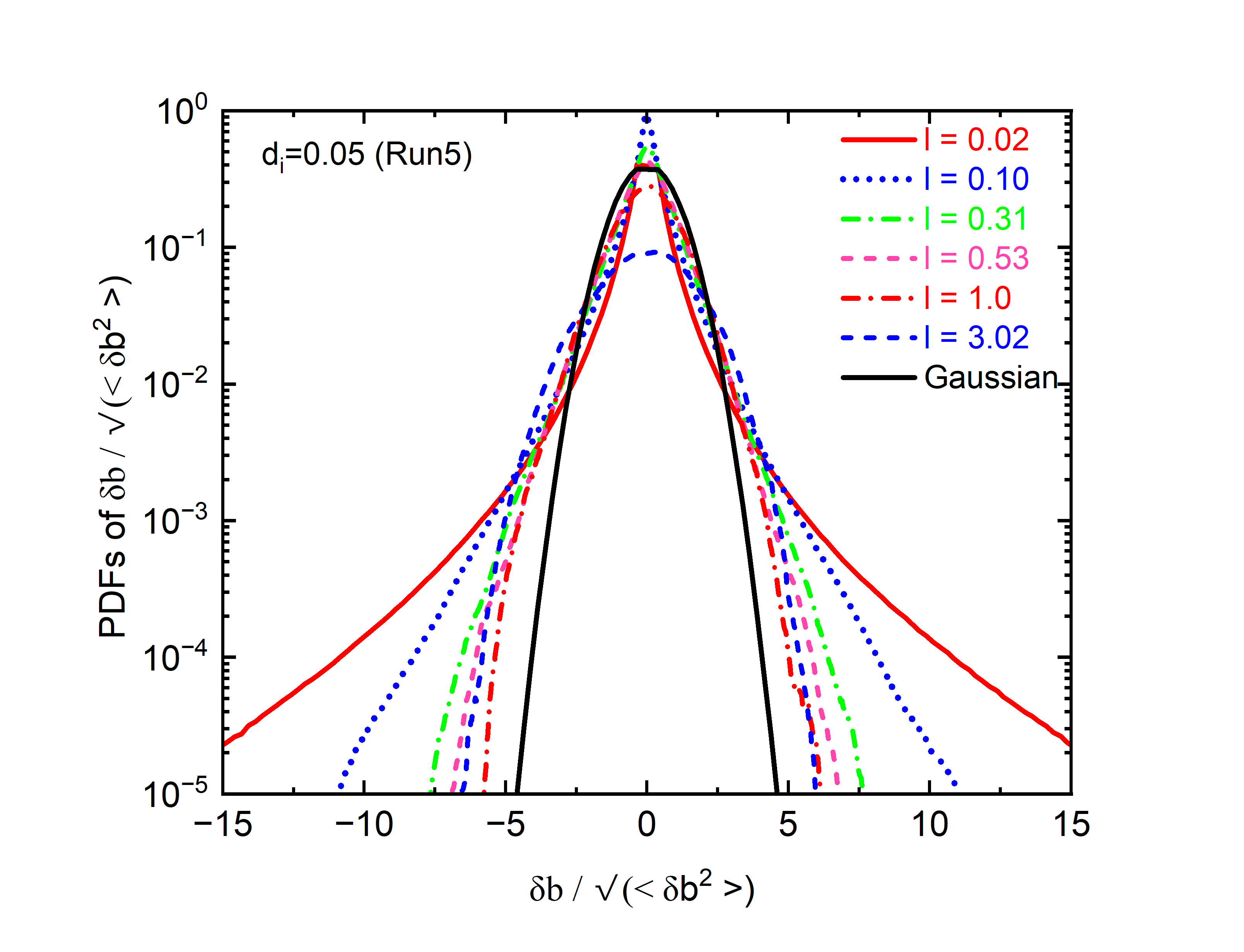} & 
\includegraphics [scale=0.22]{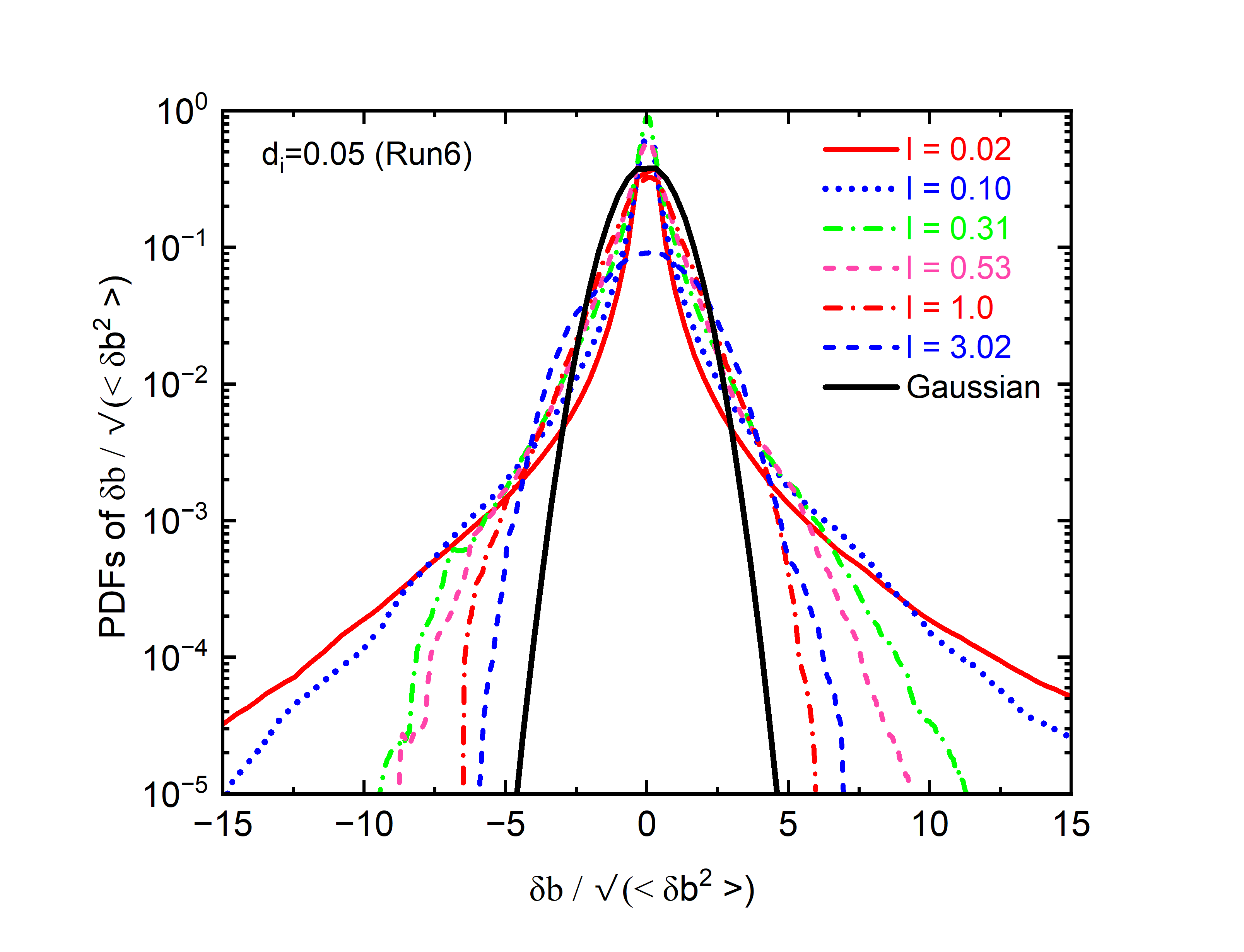}
\end{tabular}
\vskip -.1cm  
	\caption{ Semilog (base 10) plots of the PDFs of velocity field increments from (a) Run4, (b) Run5 and (c) Run6; the magnetic field increments from (d) Run4, (e) Run5, and (f) Run6 for different values of length scales $l$, which include second inertial length scales ($d_i=0.05\times2.0\pi=0.314$); for reference, we also show zero-mean and unit-variance Gaussian PDFs (black lines) in each subplot.}
    \label{fig:pdfsvelmag}
\end{figure*}

	\subsection{Energy spectra}
    \label{subsec:spectra}
In Fig.~\ref{fig:energyspectra} we display fluid- and magnetic-energy spectra, $E_u(k)$ and $E_b(k)$,  in both inertial and second-inertial (or sub-ion-scale) regions from Run4, Run5, and Run6, with $Pr_{m}=0.1,\,1$, and $10$, respectively. The log-log plots of the compensated spectra $k^{5/3}E_u(k)$ and $k^{5/3}E_b(k)$ versus the wavenumber $k$, in Figs.~\ref{fig:energyspectra} (a) and (b), are consistent with Kolmogorov-type $k^{-5/3}$ spectral scaling~\cite{frisch1995turbulence} in the first inertial region $L\ll l \ll d_{i}$ [or $2\pi/L \ll k \ll k_i \equiv 1/d_i$]; the scaling exponent is independent of $Pr_m$. 
The spectral scaling is richer in the  sub-inertial region $d_{i}\ll l \ll \eta_{d}^{b(u)}$ [see Figs.~\ref{fig:energyspectra} (c) and (d)], insofar as it depends on $Pr_m$; here, $l$ is a characteristic length scale.
Our spectra are consistent with the following sub-inertial-region scaling forms:
$E_{u}(k)\sim k^{-17/3}$ and $E_{b}(k)\sim k^{-11/3}$ [blue and black plots, for $Pr_m=1$ and $10$, respectively, in Figs.~\ref{fig:energyspectra} (c) and (d)];
and  $E_{u}(k)\sim k^{-11/3}$ and $E_{b}(k)\sim k^{-17/3}$ [red plots, for $Pr_m=0.1$, in Figs.~\ref{fig:energyspectra} (c) and (d)].

On theoretical grounds, we expect the following relations between fluid and magnetic spectra in the second inertial range:
\begin{eqnarray}
E_u(k) &\sim& k^2 E_b(k)\,,\;\; {\rm{for}}\;\; Pr_m \ll 1\,; \label{eq:prmlt1}\\
E_b(k) &\sim& k^2 E_u(k)\,,\;\; {\rm{for}}\;\; Pr_m \gg 1\,.\label{eq:prmgt1}
\end{eqnarray}
Relations~\eqref{eq:prmlt1} and \eqref{eq:prmgt1} can be obtained by the following theoretical, dominant-balance arguments. In the second inertial region, $kd_{i}\gg 1$, the energy transfer is mainly governed by the Hall term $d_{i} \nabla\times \left(\bm{J}\times \bm{b}\right)$, so the energy-transfer time $t_{r}$ can be estimated by equating the dominant terms in Eq.~\eqref{eq:mag}:
\begin{eqnarray}
\partial \bm{b}/\partial t &\sim& d_{i} \nabla\times \left(\bm{J}\times \bm{b}\right)\Rightarrow \frac{b}{t_{r}} \sim \frac{d_{i}b^{2}}{l^{2}} 
\Rightarrow t_{r}=\frac{l^{2}}{d_{i}b} \,.
\label{eq:transfer_b}
\end{eqnarray}
We first estimate the $k$-dependence of the Alfven ratio $E_{b}/E_{u}$ in low magnetic Prandtl numbers $Pr_m \ll 1$, where the kinetic energy dominates
over the magnetic energy, i.e., we expect that $\left(\bm{u}\cdot\nabla\right) \bm{u}$ plays the key role in the energy transfer in Eq. ~\eqref{eq:vel}. Therefore,
\begin{eqnarray}
\frac{\partial \bm{u}}{\partial t} &\sim & \left(\bm{u}\cdot\nabla\right) \bm{u} \Rightarrow 
\frac{u}{t_{r}} \sim \frac{u^{2}}{l} \Rightarrow
t_{r} = \frac{l}{u} 
\label{eq:transfer_v}
\end{eqnarray}
On eliminating $t_{r}$ from Eqs. ~\eqref{eq:transfer_b} and ~\eqref{eq:transfer_v}, we obtain
\begin{eqnarray}
\frac{l^{2}}{d_{i}b}= \frac{l}{u} \Rightarrow 
b^{2} = \frac{1}{d_{i}^{2}k^{2}} u^{2} \Rightarrow
\frac{E_{b}(k)}{E_{u}(k)} &=& d_{i}^{-2}k^{-2}\,,
\label{eq:loprm}
\end{eqnarray}
the result given in Eq.~\eqref{eq:prmlt1}; this arises from ion-cyclotron waves [Eq.~\eqref{eq:ionc}]. 

If $Pr_m \gg 1$, the magnetic energy dominates over its fluid counterpart, so the term
$\left(\bm{b}\cdot\nabla\right) \bm{b}$ in Eq.~\eqref{eq:vel} plays the key role in energy transfer,
therefore,
\begin{eqnarray}
\frac{\partial \bm{u}}{\partial t} &\sim & \left(\bm{b}\cdot\nabla\right) \bm{b} \Rightarrow 
\frac{u}{t_{r}} \sim \frac{b^{2}}{l} \Rightarrow
t_{r} = \frac{lu}{b^{2}}\,.
\label{eq:transfer_vb}
\end{eqnarray}
We can now eliminate $t_{r}$ from Eqs.~\eqref{eq:transfer_b} and ~\eqref{eq:transfer_vb} to get
\begin{eqnarray}
\frac{lu}{b^{2}}= \frac{l^{2}}{d_{i}b} \Rightarrow 
b^{2} = d_{i}^{2}k^{2} u^{2} \Rightarrow
\frac{E_{b}}{E_{u}} &=& d_{i}^{2}k^{2}\,,
\label{eq:hiprm}
\end{eqnarray}
the result given in Eq.~\eqref{eq:prmgt1}; this arises from whistler waves [Eq.~\eqref{eq:whis}].  

The Alfv\'en ratios~\eqref{eq:loprm} and ~\eqref{eq:hiprm} allow us to relate the scaling exponents of the power-law 
regimes in fluid and magnetic-energy spectra, $E_{u}(k)$ and $E_{b}(k)$, respectively.
Our DNS results, for $Pr_{m}=1$ and $10$ [Runs 5 and 6], show that, in sub-inertial region, $E_{u}(k)\sim k^{-17/3}$ [blue and black plots in Fig.~\ref{fig:energyspectra} (c)] and  $E_{b}(k)\sim k^{-11/3}$ [blue and black plots in Fig.~\ref{fig:energyspectra} (d)], which is consistent with the large-$Pr_{m}$ Alfven ratio~\eqref{eq:hiprm}
for the whistler mode. Similarly, from the red curves in Figs.~\ref{fig:energyspectra} (c) and (d) for $Pr_{m}=0.1$ [Run 4],
we see that, in in the sub-inertial region, $E_{u}(k)\sim k^{-11/3}$ and $E_{b}(k)\sim k^{-17/3}$, which is consistent with the low-$Pr_{m}$ Alfv\'en ratio~\eqref{eq:loprm} for the ion-cyclotron mode. In Fig.~\eqref{fig:alfvenratiospectra} we display log-log plots of the Alfv\'en ratio $E_{b}\left(k\right)/E_{u}\left(k\right)$ versus $k$ for runs Run$1$-Run$8$ [see Table~\ref{table:1}]; these plots show regions that are consistent with $k^2$ and $k^{-2}$ power-law ranges [cf. Eqs.~\eqref{eq:hiprm} and \eqref{eq:loprm}]], even though the ranges of scales available in our DNSs are limited.

We turn now to an examination of $E_R(k)$ and $E_L(k)$, the energy spectra~\eqref{eq:rightleftfluct} for which the polarization ${\mathcal{P}}_m(k) < 0$ and ${\mathcal{P}}_m(k) > 0$, respectively. The upper panels in Figs.~\ref{fig:fluctuationrun456}  and ~\ref{fig:fluctuationrun78} display, respectively, plots of $E_{R}(k)/E_{T}(k)$ [red] and $E_{L}(k)/E_{T}(k)$ [blue] versus $k$ for Run$4$-Run$6$ and Run$7$-Run$8$, which show that, if $Pr_{m}\geq1$, then $E_{R}(k) > E_{L}(k)$ for $d_i = 0.05$ [Fig.~\ref{fig:fluctuationrun456}] and 
$d_i = 0.1$ [Figs.~\ref{fig:fluctuationrun78} (b) and (c)], i.e., $R$ fluctuations dominate over their $L$ counterparts and this dominance increases with $k$. For $Pr_{m}=0.1$, $E_{L}(k) > E_{R}(k)$ for  $d_i = 0.1$ [Figs.~\ref{fig:fluctuationrun78} (a)]; 
for  $d_i = 0.05$ [Figs.~\ref{fig:fluctuationrun456} (a)], the difference between $E_R(k)$ and $E_L(k)$ is reduced relative to that in the plots for $Pr_{m}\geq1$ in Figs.~\ref{fig:fluctuationrun456} (b) and (c); i.e., if $Pr_{m}\leq1$, then $L$ fluctuations begin to dominate over their $R$ counterparts. The precise value of $Pr_m$ at which $R$ dominance gives way to $L$ dominance depends on $d_i$ and the resolution of the runs. [We present such plots for Run1, Run2, and Run3 in the Appendix.]

The lower panels of Figs.~\eqref{fig:fluctuationrun456} and ~\eqref{fig:fluctuationrun78} contain log-log plots of the compensated magnetic energy spectra, for both $L-$ and $R-$ fluctuations. These plots indicate that, if $Pr_{m}\ll1$,  then the magnetic energy spectra of both $L-$ and $R-$ fluctuations show power-law ranges $\sim k^{-17/3}$\, whereas, if $Pr_{m}\geq1$, then they show power-law ranges $\sim k^{-11/3}$ in both fluctuations. Again, the precise value of $Pr_{m}$, for the crossover from one of these ranges to the other one, depends on $d_i$ and the resolution of the runs.

\subsection{Intermittency}
\label{subsec:inter}
We now investigate intermittency in decaying 3D HMHD turbulence, at the cascade-completion time $t_c$, by quantifying  the 
dependence, on the length scale $l$, of the PDFs of the longitudinal-field increments $\delta a_{\parallel}$, the structure functions $S^{a}_{p}$, and the flatnesses $F_{4}^{a}$ [Eq.~\eqref{eq:statproperties}].


The PDFs of velocity- and magnetic-field increments are shown in Fig.~\ref{fig:pdfsvelmag} for Run$4$-Run$6$ and different values of $l$ that span the first and second inertial scales and also the dissipation regions [$l= 0.02$ (red-solid curve), $l=0.10$ (blue-dotted-curve), $l=0.31$ (green-dash-dot-curve), $l=0.53$ (pink-dash-curve), $l=1.0$ (red-dash-dot-curve), and $l= 3.02$ (blue-dash-curve). For comparison Gaussian PDFs are shown by black curves. The top (bottom) panel of Fig.~\ref{fig:pdfsvelmag} portrays PDFs of velocity (magnetic-field) increments. The scale-dependent heavy tails of these PDFs show clear deviations from Gaussian statistics and provide the first signatures of intermittency. These PDFs also depend upon $Pr_m$. The velocity-increment PDFs of velocity increments show that intermittency is suppressed as $Pr_m$ increases; by contrast, magnetic-field-increment PDFs show the opposite trend, i.e., intermittency increases  with $Pr_m$.
At large $l$, comparable to the initial energy-injection length scales, both these PDFs are nearly Gaussian for all the values of $Pr_m$ that we have considered.

We quantify this intermittency by examining the order-$p$ longitudinal structure functions and associated flatnesses~\eqref{eq:statproperties}, 
    Any multiscaling, in either the first or the second inertial regions, can be charaterized by obtaining the 
    multiscaling exponents that are defined as follows:
    \begin{eqnarray}
    S_{p}^{u}\left(l\right) &\sim& l^{\xi_{p}^{u}}\,; \nonumber \\ 
    S_{p}^{b}\left(l\right)&\sim& l^{\xi_{p}^{b_{1}}\left(\xi_{p}^{b_{2}}\right)}\,;
    \end{eqnarray}
    here, $\xi_{p}^{u}$ and $\xi_{p}^{b_{1}}$ are, respectively, the order-$p$ multiscaling exponents for velocity- and magnetic-field
    increments in the first inertial region $L\ll l \ll d_{i}$; and $\xi_{p}^{b_{2}}$ is the order-$p$ multiscaling exponent of the magnetic-field increment in the second inertial region $d_{i}\ll l \ll \eta_{d}^{b}$. These exponents are linear functions of $p$ if the turbulent state exhibits simple scaling, but they are nonlinear functions of $p$ when it displays 
    multiscaling~\cite{frisch1995turbulence}. If experimental or computational constraints limit the spatial resolution that is available, we can employ  
    the extended self-similarity (ESS) procedure~\cite{yadav2022statistical,benzi1993extended} fruitfully to determine the ratios of multiscaling exponent from the slopes of log-log plots of one structure function versus another [say, $S_p(l)$ versus $S_3(l)$] in the appropriate inertial region. We use this ESS procedure given the challenges that are inherent in DNSs of 3D HMHD, especially when $Pr_m$ is different from unity.
    
\begin{figure*}[t]
\centering 
\begin{tabular}{c c c}
\text{(a)} & \text{(b)} & \text{(c)}  \\
\includegraphics [scale=0.22]{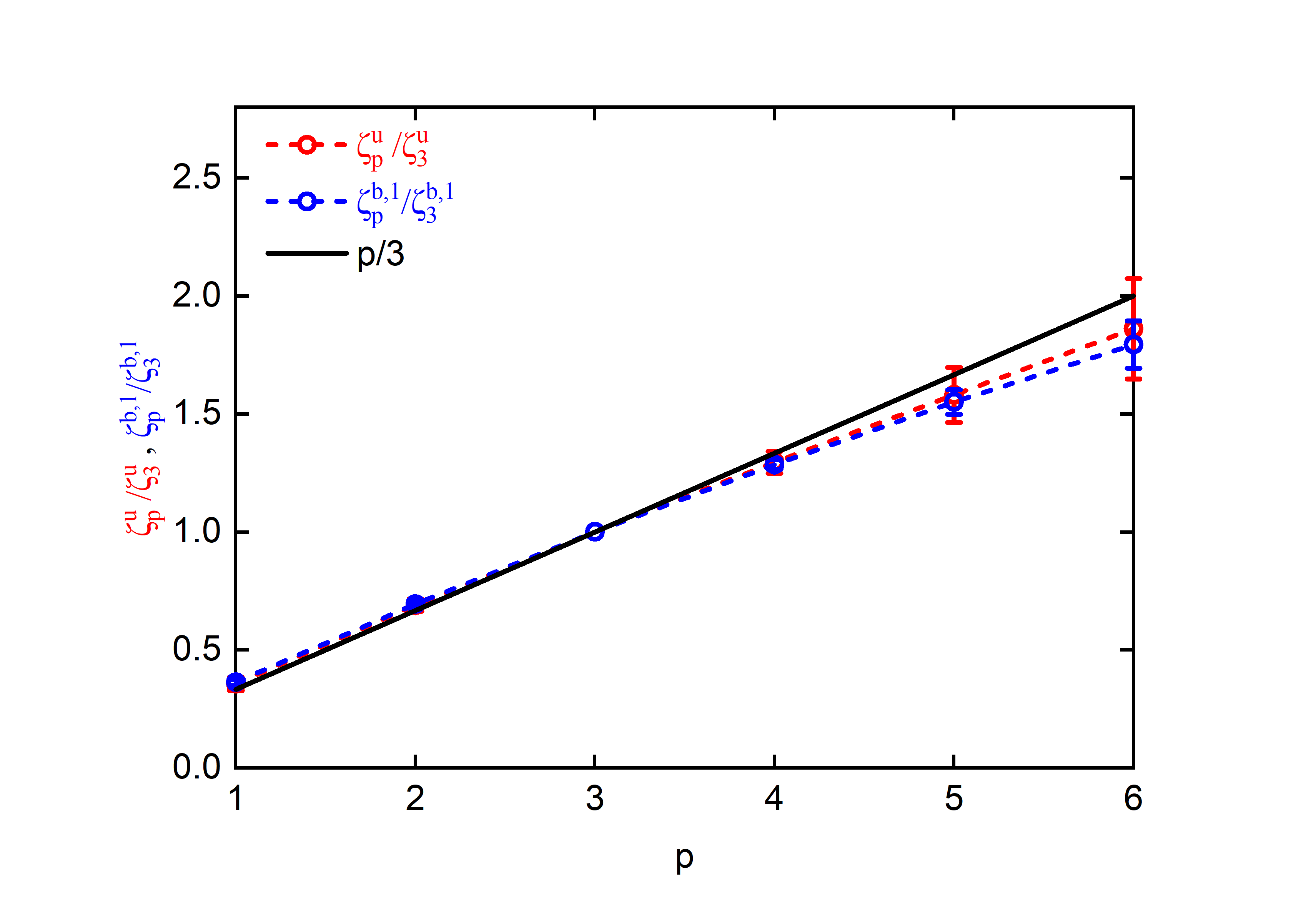} & 
\includegraphics [scale=0.22]{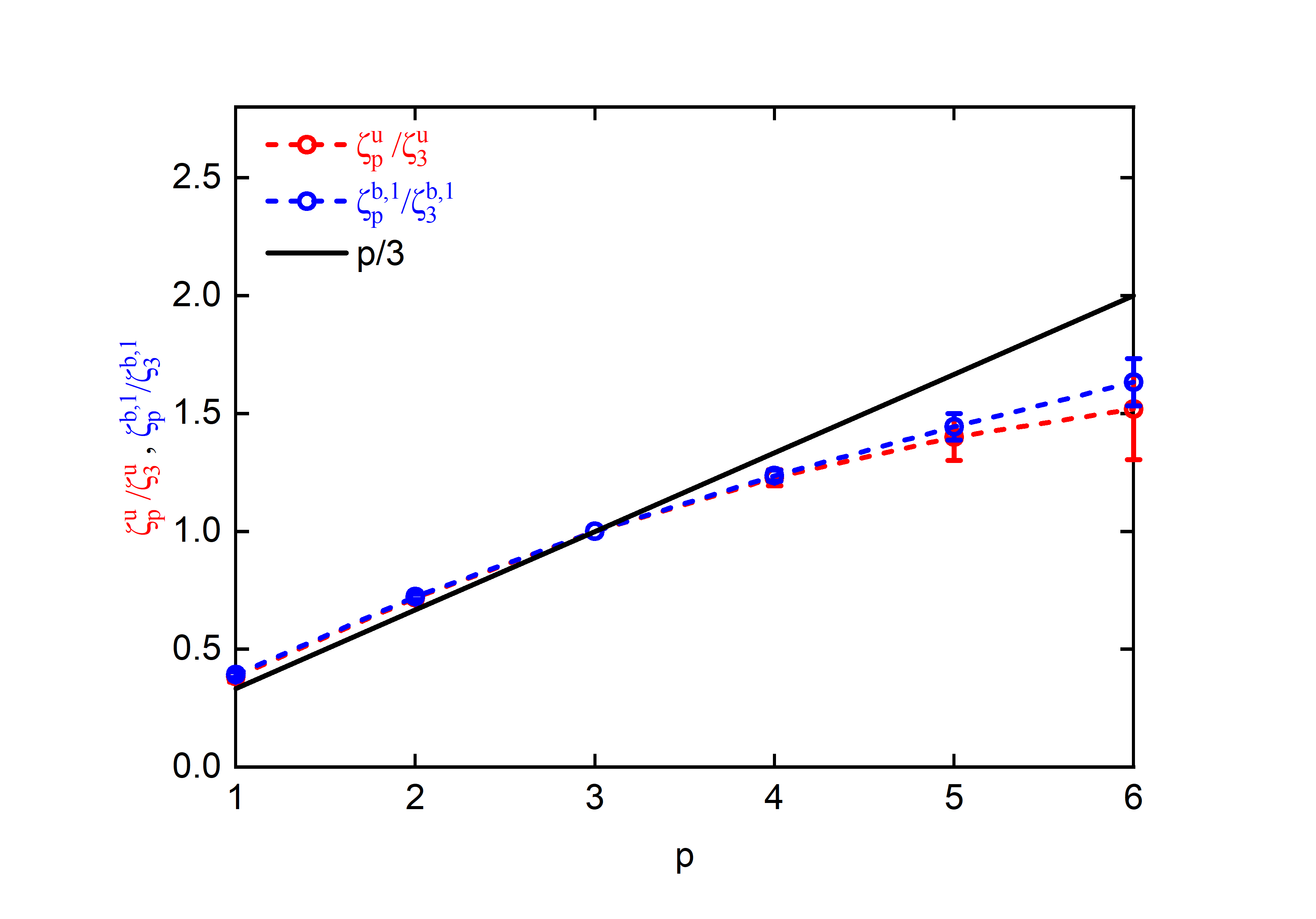} &
\includegraphics [scale=0.22]{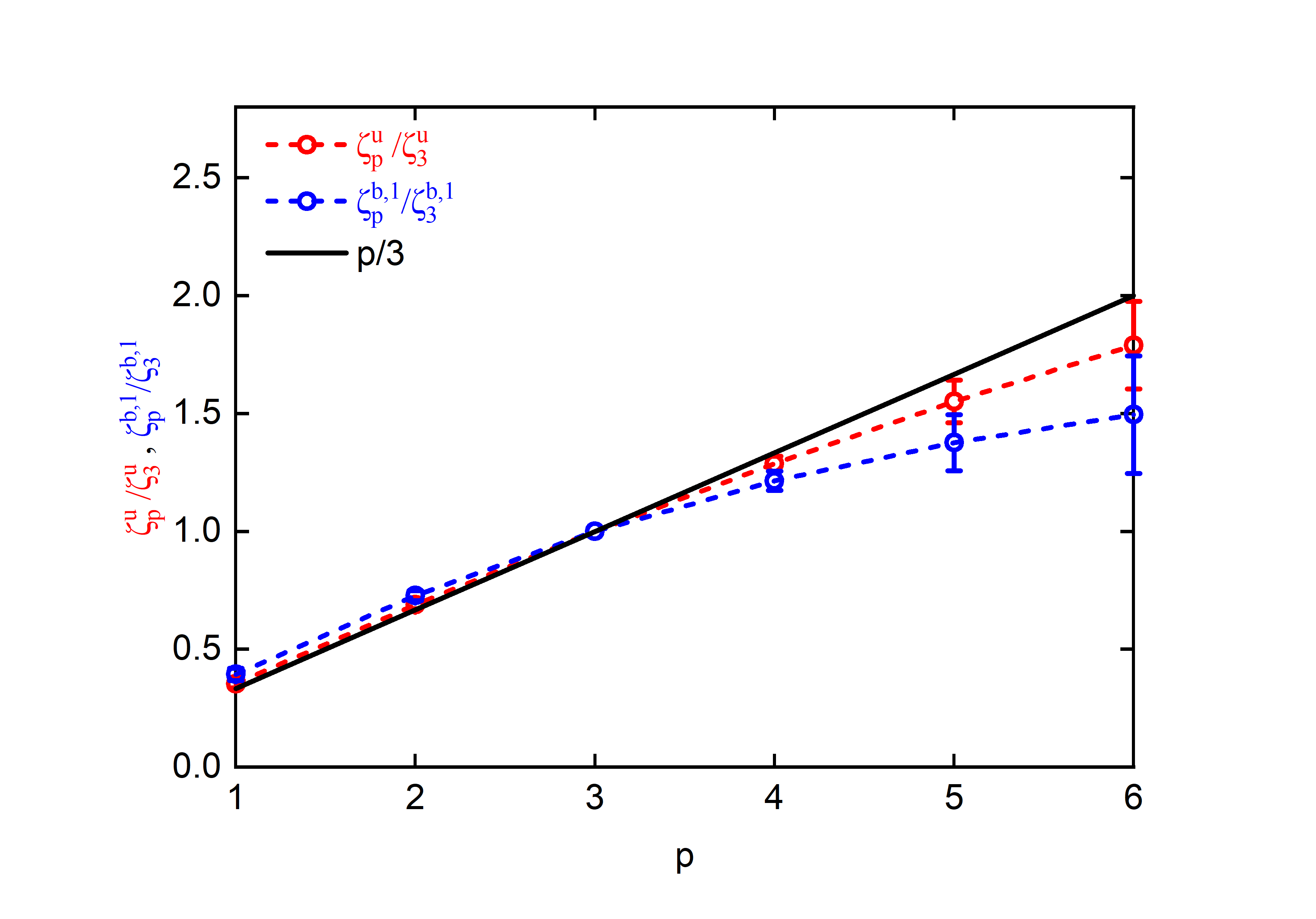}
\end{tabular}
%
\begin{tabular}{c c c}
\text{(d)} & \text{(e)} & \text{(f)}  \\
\includegraphics [scale=0.22]{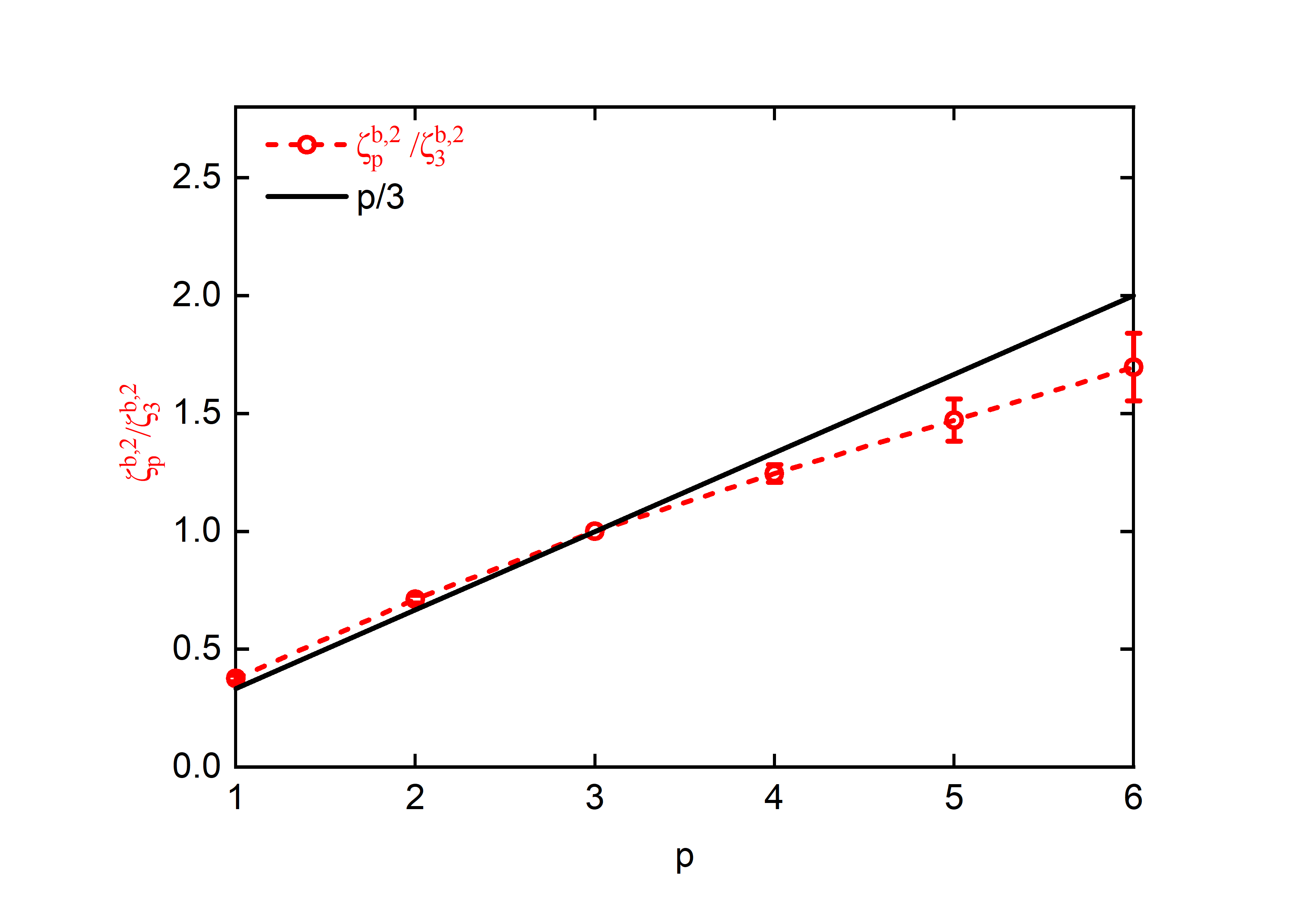} &
\includegraphics [scale=0.22]{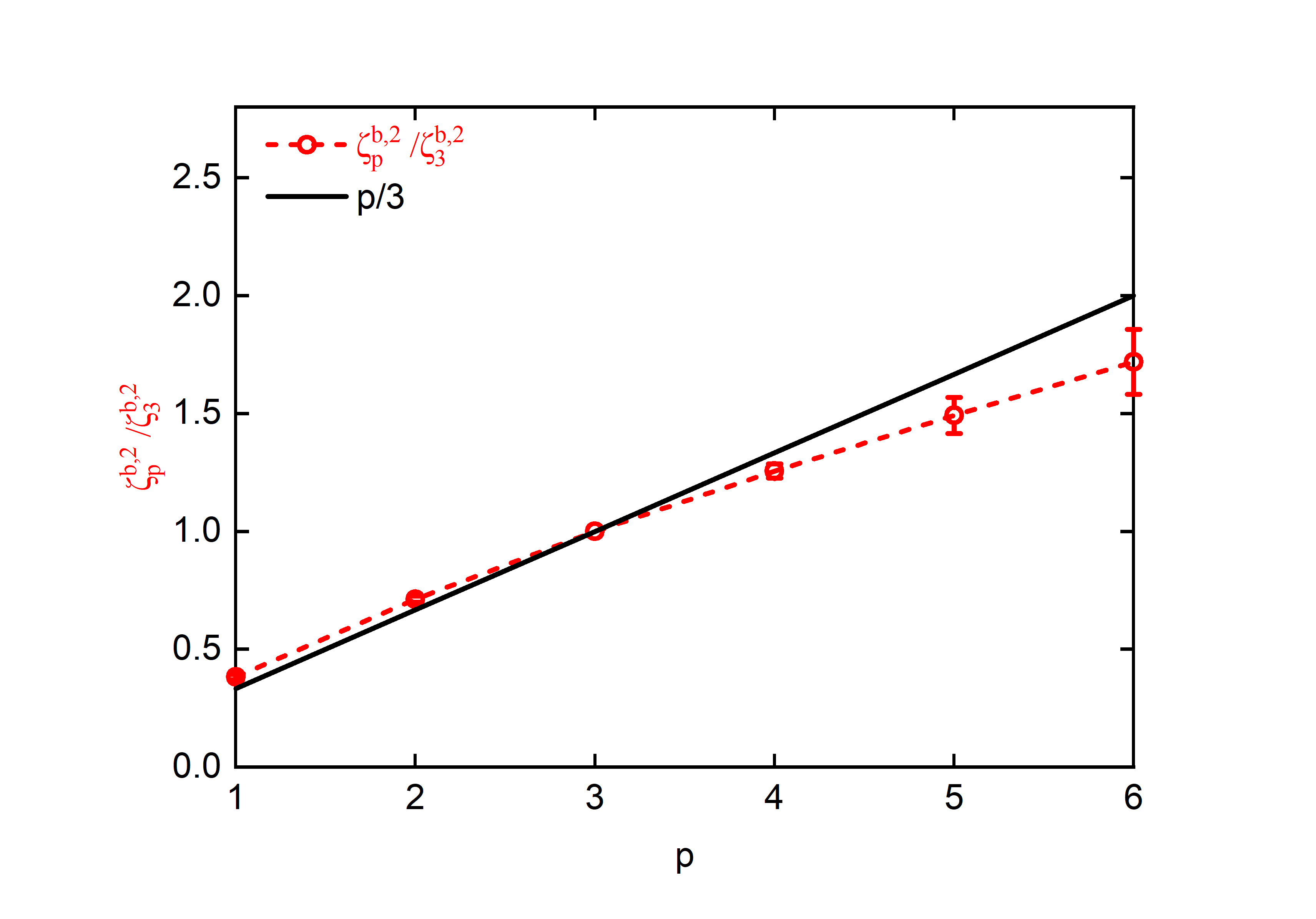} & 
\includegraphics [scale=0.22]{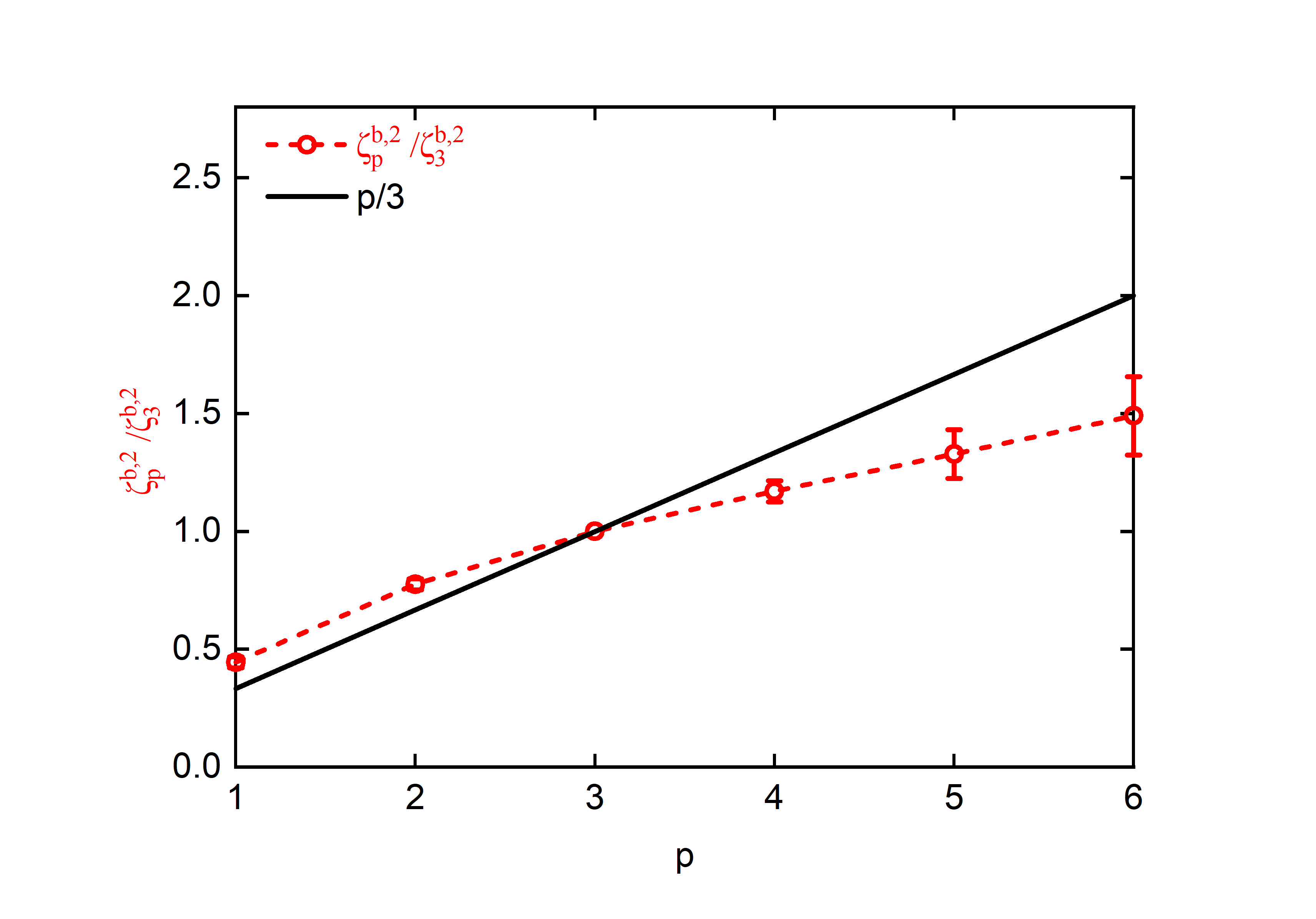}
\end{tabular}
\vskip -.1cm

\caption{ Plots of exponent ratios {\it{vs.}} order $p-$ for velocity and magnetic fields in inertial $\left(\zeta_p^u(l)/\zeta_3^u(l), \zeta_p^{b,1}(l)/\zeta_3^{b,1}(l)\right)$ (top-panel) and second-inertial  $\left(\zeta_p^{b,2}(l)/\zeta_3^{b,2}(l)\right)$ (bottom-panel) regions from simulations (a,d) Run4, (b,e) Run5 and (c,f) Run6; for comparison in all subplots we also display the Kolmogrov variation of exponents ($p/3$) with order $p-$ shown by black curves; the exponent ratios shown here are obtained from the ESS plots of the corresponding structure functions shown in appendix. for Run4, Run5 and Run6 $Pr_{m}=0.1$,$1.0$ and $10.0$ respectively }

\label{fig:multiexponentrun456}
\end{figure*}

\begin{figure*}[t]
\centering 
\begin{tabular}{c c c}
\text{(a)} & \text{(b)} & \text{(c)}  \\
\includegraphics [scale=0.23]{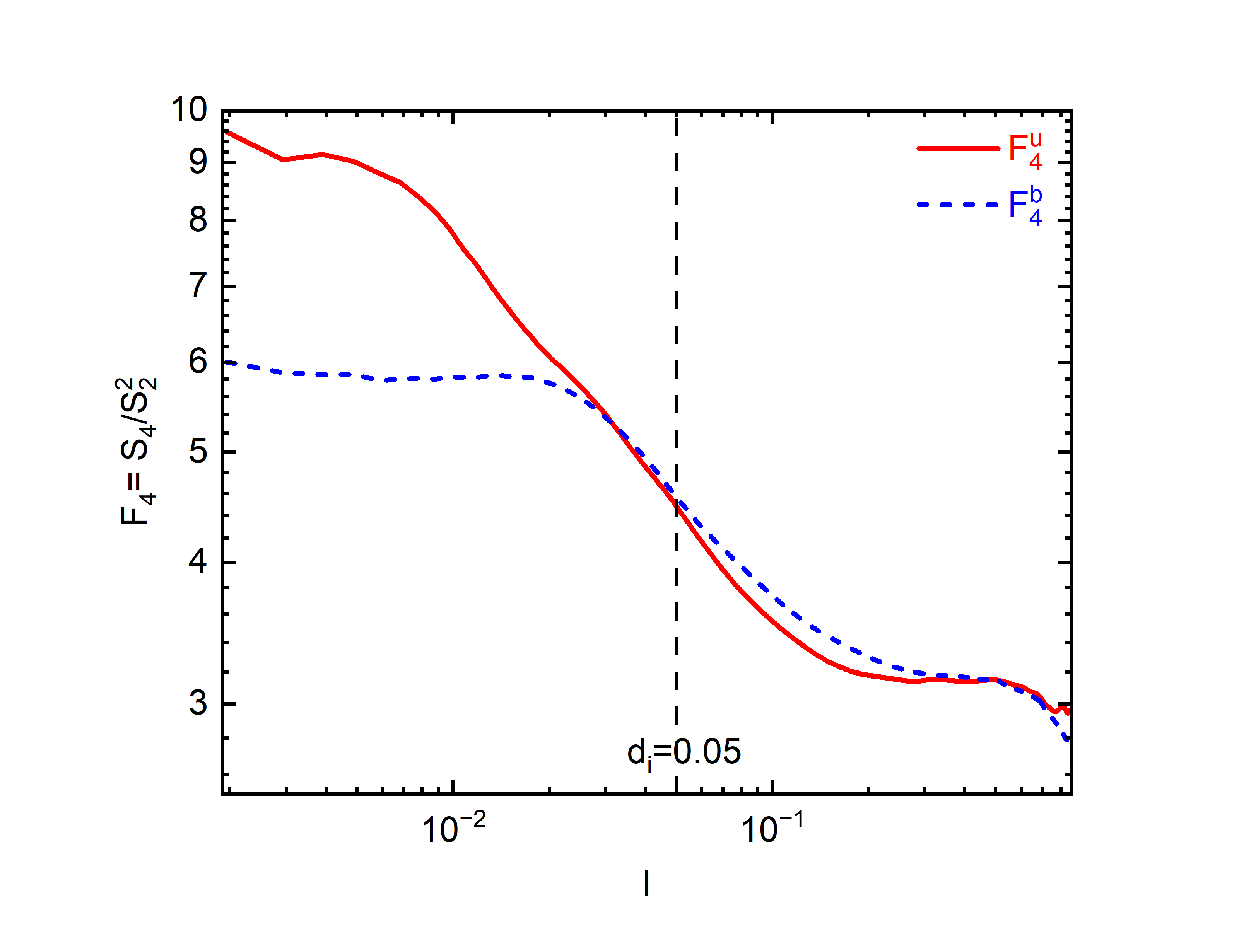} & 
\includegraphics [scale=0.23]{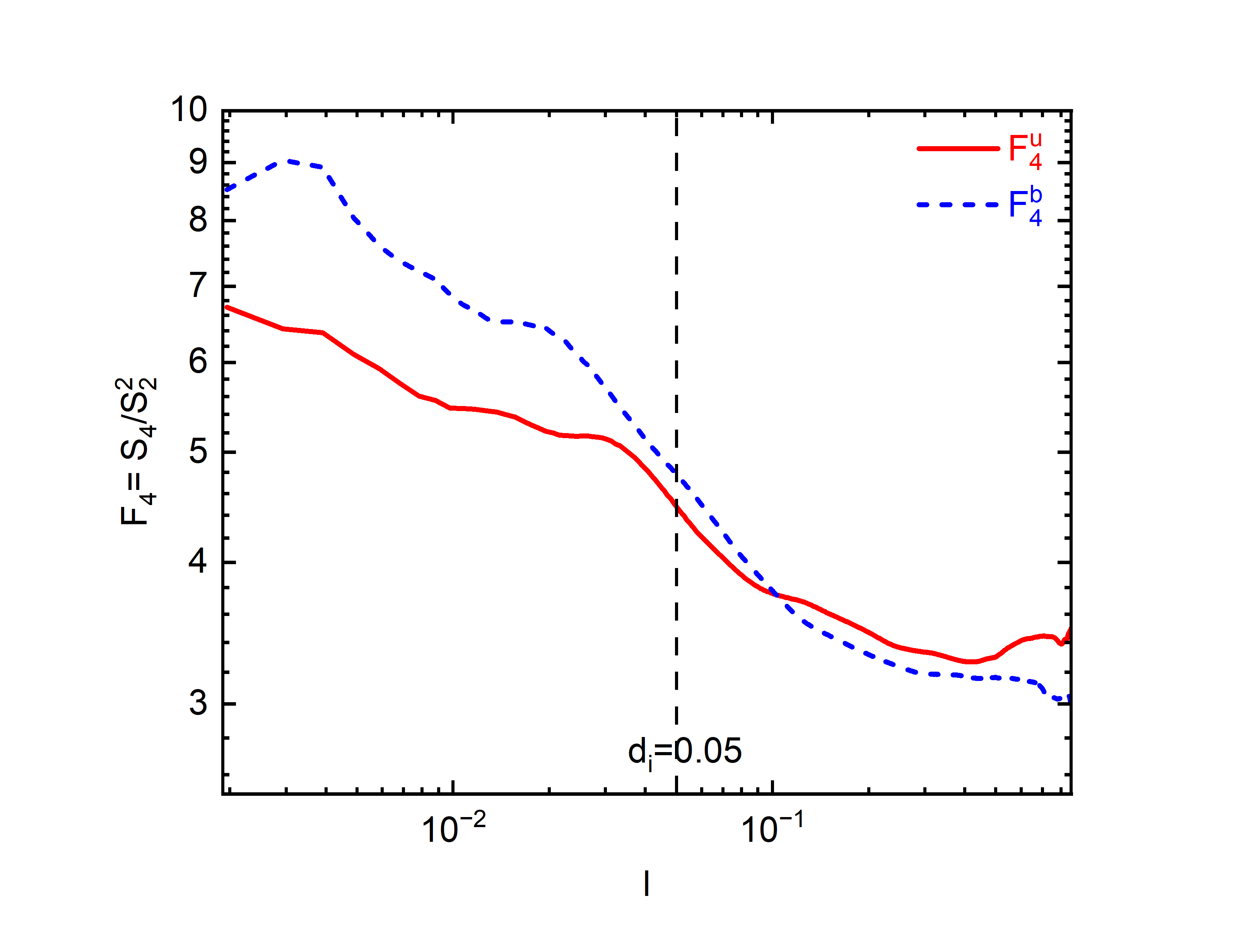} &
\includegraphics [scale=0.23]{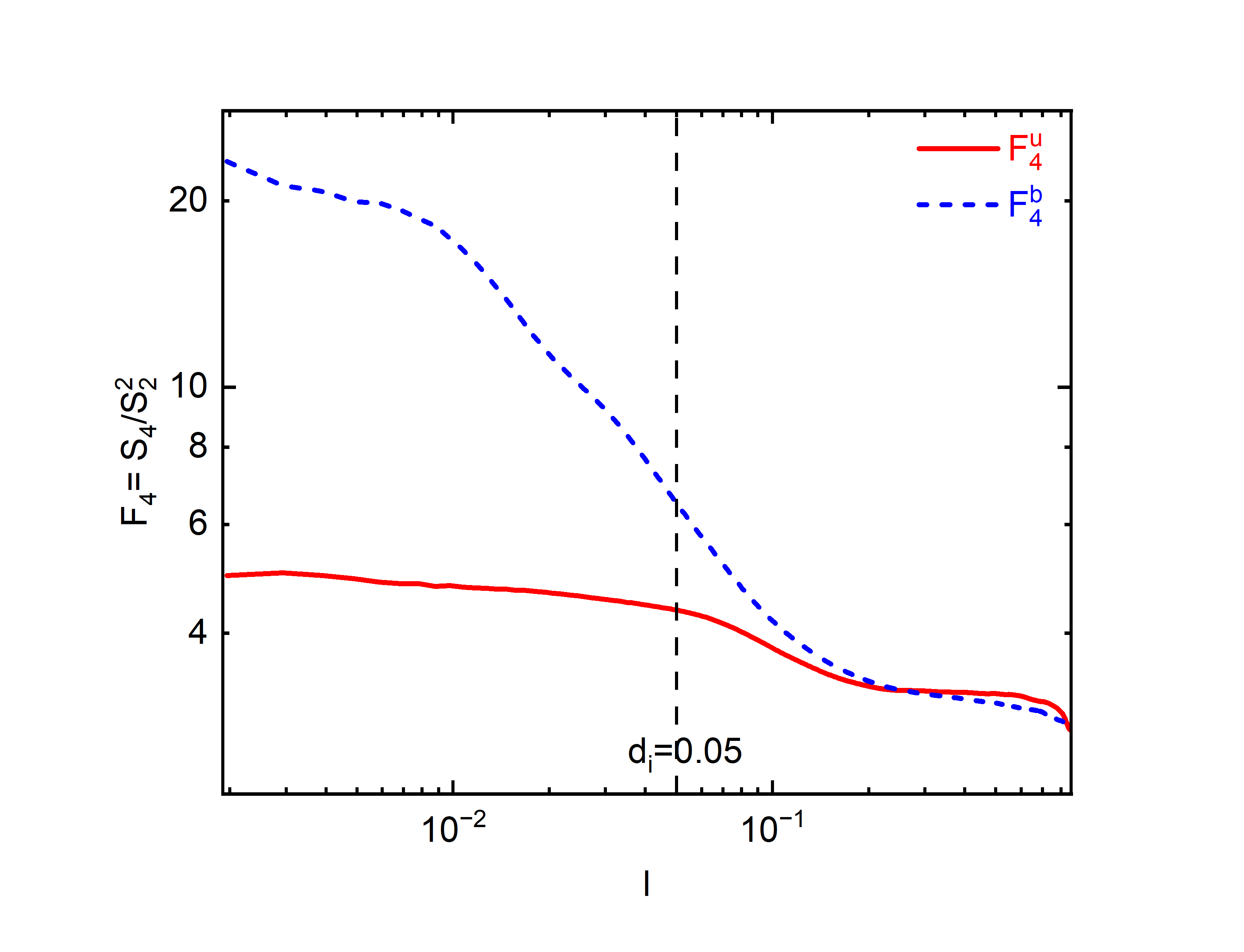}
\end{tabular}
\vskip -0.1cm 
	\caption{
    Log-log (base 10) plots of the flatnesses for the velocity $F_4^\text{u}$ (red solid line) and magnetic-field $F_4^\text{b}$ (blue dashed line) for (a) Run4, (b) Run5 and (d) Run6; Run4, Run5 and Run6 are the HMHD simulations, respectively, at $Pr_{m}=0.1$, $1.0$ and $10.0$.}
    \label{fig:flatnessrun456}
\end{figure*}

    Figure~\ref{fig:multiexponentrun456} displays plots of the multiscaling exponent ratios vs the order $p$ for both velocity- and magnetic-field increments from Run4 [Figs.~\ref{fig:multiexponentrun456}(a) and (d)], Run5 [Figs.~\ref{fig:multiexponentrun456}(b) and (e)], and Run6 [Figs.~\ref{fig:multiexponentrun456}(c) and (f)]. The top panel of Fig.~\ref{fig:multiexponentrun456} presents
    plots of $\zeta_{p}^{u}/\zeta_{3}^{u}$ and $\zeta_{p}^{b,1}/\zeta_{3}^{b,1}$ in the first inertial region, whereas the bottom panel shows $\zeta_{p}^{b,2}/\zeta_{3}^{b,2}$ for the magnetic-field increments in the second-inertial region.
    For comparison, the linear Kolmogorov-type [K41] prediction for these exponents is shown by the black straight lines;
    deviations of the multiscaling exponent ratios from these straight lines are a measure of the intermittency.
    For all our runs, this intermittency is significant for $ p  > 3$. In the second inertial range, these deviations 
    increase with $Pr_m$ and are more marked for $Pr_{m}=10.0$ than for $Pr_{m}=0.1$ .
    Intermittency can also be studied examining the scale-dependence of the ehavior of flatnesses for velocity- and magnetic-field increments [see Eq. ~\eqref{eq:statproperties}]. 
    In Fig.~\eqref{fig:flatnessrun456} we display the plots versus $l$ of the flatness for the increments of the velocity (red-solid-curve) and the magnetic field (blue-dashed-curve) for simulations Run4 [Fig.~\eqref{fig:flatnessrun456} (a)],
    Run5 [Fig.~\eqref{fig:flatnessrun456} (b)], and Run6 [Fig.~\eqref{fig:flatnessrun456} (c)].
    For large length scales ($l >d_{i}$), these flatnesses are comparable to the Gaussian value of $3$. By contrast,
    at small length scales ($l \le d_{i}$) these flatnesess increase significantly, and rise well above the Gaussian value of $3$, for all values of  $Pr_{m}$ that we consider; this is a clear indication of small-scale intermittency in 3D HMHD turbulence. 
    The velocity flatnessess increases faster than its magnetic-field counterpart when $Pr_{m}=0.1$; whereas, if $Pr_{m}=1.0$ and $10.0$, the roles are reversed and magnetic flatness increases more rapidly than the velocity flatness.
    Like other analysis, the nature of flatness also supports our claim that the magnetic fields are more intermittent in comparison to its velocity counterpart at smaller length scales for $Pr_{m}=1.0$ and $10.0$. \par

    We have presented detailed results from our DNSs Run4, Run5, and Run6. Our results from Run1-Run3 and Run7 and Run8 are similar, so we have not presented them in detail here.

\section{Conclusions}
\label{sec:conc}

The 3D HMHD equations are of relevance in many physical settings~\cite{Huba2003,shaikh20093d,matthaeus2012review,miura2024formation}, like the  the magneto-rotational instability~\cite{o2014multifluid,kopp2022influence}, the solar wind~\cite{matthaeus1982measurement,krishan1999astrophysical,matthaeus2012review,kiyani2009global}, magnetic-reconnection processes~\cite{yamada2010magnetic}, pulsars and neutron stars~\cite{shalybkov1997hall,hollerbach2002influence,igoshev2021evolution}, magnetic-field transport in plasma opening switches
, and 
sub-Alfv\'enic plasma expansion~\cite{ripin1993sub}. In particular, the development of an understanding of the statistical properties of solar-wind turbulence continues to be a major challenge in solar plasma physics because satellite observations of solar-wind-plasma
turbulence yield a \textit{fluid-energy spectrum} $E_u(k)$, with $k$ the wavenumber, with a power-law inertial range that has a scaling exponent that is consistent with $-5/3$ [the Kolmogorov prediction (henceforth K41)] and a \textit{magnetic-energy spectrum} $E_b(k)$ that displays (i) a first inertial range, consistent with K41, and (ii) a second inertial range that is characterized by an exponent $\in [-4,-1]$. These observations also uncover  intermittency and multiscaling of velocity and magnetic-ﬁeld structure functions  in the first inertial range, but not in the second one. It is important, therefore, to explore such scaling regions and intermittency in 3D HMHD. Our DNSs have been designed to study systematically these statistical properties as a function of two important control parameters, namely, $d_i$ and $Pr_{m}$ for unforced 3D HMHD. This has never 
been attempted. At $Pr_m=1$ these DNSs yield~\cite{yadav2022statistical,miura2024formation} a K41-type inertial range for $E_u(k)$ spectrum  and $-7/3$ and $-11/3$ scaling regimes in $E_b(k)$ in the sub-inertial region, with $d_{i} \ll l \ll \eta_{d}^{b}$, where $d_{i}$ is the ion-inertial length and $\eta_{d}^{b}$ the magnetic-energy dissipation length scale. There are very few studies of such scaling and intermittency in 3D HMHD turbulence when $Pr_{m}\neq1$. We first show on theoretical grounds that, in the second inertial range, $E_u(k) \sim k^2 E_b(k)$, for $Pr_m \ll 1$ and  $E_b(k) \sim k^2 E_u(k)$, for $Pr_m \gg 1$. We then carry out direct numerical simulations (DNSs) that uncover such 
scaling forms. We present a detailed study of such turbulence at values of $Pr_{m}$ that are different from unity. We compare our results for $Pr_{m}=0.1$, $Pr_{m}=1$, and $Pr_{m}=10$, at  
 different values of $d_{i}$, and with $512^3$ and $1024^3$ collocation points in our pseudospectral DNSs. We compute various statistical quantities to characterize 3D HMHD plasma turbulence:
 (i) $E_u(k)$ and $E_b(k)$, including spectra conditioned on left- and right-polarised fluctuations of the fields, to uncover ion-cyclotron and whistler waves in the sub-inertial region; (ii) probability distribution functions (PDFs) of longitudinal velocity and magnetic-fields increments, their structure functions, and their flatnesses.  For all values of $Pr_{m}$, there is a K41-type inertial range with  $E_u(k) \sim k^{-5/3}$ and
  $E_b(k) \sim k^{-5/3}$. For large $k$, i.e., length scales smaller than $d_i$, we find that $E_b(k) \sim k^{-17/3}$ for $Pr_{m}=0.1$; but $E_b(k) \sim k^{-17/3}$ for  $Pr_{m}=1$ and  $Pr_{m}=10$. We present theoretical arguments for these two different scaling behaviors of $E_b(k)$, in this second-inertial region, and show that they can be understood using left- and right-polarised fluctuations of the fields, which lead to the dominance of either ion-cyclotron or whistler waves. Our analysis of PDFs of field increments, structure functions, and their flatnesses show that the velocity fields are less intermittent than their magnetic-field counterparts for  $Pr_{m}=1$ and  $Pr_{m}=10$. The specific predictions we make for these statistical properties of 3D HMHD turbulence should be measured in all physical realisations of the 3D HMHD system.

\begin{acknowledgments}
SY and RP thank the Anusandhan National Research Foundation (ANRF), the Science and Engineering Research Board (SERB), and the National Supercomputing Mission (NSM), India, for support,  the Supercomputer Education and Research Centre (IISc), for computational resources, and NIFS (Japan) for hospitality while part of this work was being done. The numerical simulations were performed on an NEC SX-Aurora TSUBASA A412-8 Plasma Simulator at the National Institute of Fusion Science (NIFS), Japan, with the support and under the auspices of the NIFS Collaboration Research program (NIFS23KISS030), FUJITSU FX1000 Wisteria/BDEC-01 Odyssey of the University of Tokyo with partial support from the “Joint Usage/Research Center for Interdisciplinary Large-scale Information Infrastructures” (Project ID:jh230004, jh240004), and FUJITSU/RIKEN FUGAKU with the support of the HPCI System Research Project (Project ID: hp240026) in Japan.
This research was partially supported by JSPS KAKENHI Grant Numbers 24K06893, of Japan, and by the NINS program of Promoting Research by Networking among Institutions (Grant Number 01422301).
This work was supported by the Research Institute for Mathematical Sciences, an International Joint Usage/Research Center located in Kyoto University.
\end{acknowledgments}
\bibliography{mybib}
\end{document}